\documentclass[%
reprint,
 amsmath,amssymb,
 aps,
pra,
floatfix,
longbibliography,
]{revtex4-1}

\usepackage{graphicx}
\usepackage{dcolumn}
\usepackage{bm}

\usepackage{multirow}
\usepackage{array}
\usepackage{color}

\usepackage{float}

\begin{document}

\preprint{APS/123-QED}

\title{Toroidic and antitoroidic orders in hexagonal arrays of dielectric trimers: \\
magnetic group approach}

\author{Victor~Dmitriev$^{1}$}
\author{Silvio~Domingos~Silva~Santos$^{1}$}
\author{Andrey~B.~Evlyukhin$^{2}$}
\author{Anton~S.~Kupriianov$^{3}$}
\author{Vladimir~R.~Tuz$^{4}$}
\email{tvr@jlu.edu.cn}
\affiliation{$^1$Electrical Engineering Department, Federal University of Para, R. Augusto Correa 01, CEP 66075-900 Belem, Para, Brazil}
\affiliation{$^2$Institute of Quantum Optics, Leibniz Universit{\"a}t Hannover, Welfengarten Street 1, 30167 Hannover,  Germany}
\affiliation{$^3$College of Physics, Jilin University, 2699 Qianjin Street, Changchun 130012, China}
\affiliation{$^4$State Key Laboratory of Integrated Optoelectronics, College of Electronic Science and Engineering, International Center of Future Science, Jilin University, 2699 Qianjin Street, Changchun 130012, China}

\date{\today}

\begin{abstract}
Herein, we investigate  symmetry-protected toroidal dipole resonances and conditions of their excitation in a new type of electromagnetic metamaterials. These metamaterials are all-dielectric planar periodic arrays of dielectric disks disposed on a dielectric substrate. The elementary building blocks of the array are trimers which are distributed in hexagonal unit supercells. The highest geometrical symmetry of the unit supercell is $C_{6v}$. The analysis is fulfilled by using the representation theory of groups with  application of the magnetic group theory, which is a new approach in solving such problems. We have shown that to get access to the toroidal supermodes of the array,  symmetry of  the unit supercell must be broken twice: firstly, the $C_{3v}$ symmetry of the trimer, and secondly, the $C_{6v}$ symmetry of the unit supercell needs to be reduced. Selection rules for the symmetric and antisymmetric orders of the toroidal dipole moments in the arrays are defined. In particular, we have shown that with the reduction of the unit supercell symmetry to the $C_{2v}$ group, the array exhibits the toroidal dipole resonance with  antitoroidic order. The arrays with the lower $C_{s}$ symmetry can provide the resonances with both toroidic and antitoroidic orders. It is also shown that these arrays are always polarization sensitive. Full-wave simulations and experiments confirm the theoretical predictions. The suggested metamaterials can provide an enhanced light-matter interaction due to the spatially and temporally confined light in resonant systems with very high quality factors. 
\end{abstract}


\maketitle

\section{Introduction}
The majority of physical systems possess intrinsic symmetries which define the areas of possible solutions of the equations governing these systems. With the use of group-theoretical methods, one can classify the corresponding solutions defined by the underlying symmetries. In particular, symmetry analysis by the group-theoretical approach has been successfully applied to simplify the description of different physical phenomena and used as a general guideline to design many practical devices in electronics, acoustics, and optics \cite{Rousseau_raman_1981, Bak_PhysRevB_1985, Sakoda_PhysRevB_1995, Ochiai_PhysRevB_2001, Hergert_2003, Mock_PhysRevB_2010, Alagappan_PhysRevB_2008}. To date, there exist many textbooks describing the use of group theory for solving specific physical problems \cite{hamermesh_book_1962, Tinkham_book_1964, cornwell_book_1984, Ludwig_book_1988, Inui_book_1996, Bradley_book_2009, hergert_book_2018}.

Besides simplification of numerical calculations, group theory can be applied to classify promising systems for further investigations, such as in the case of  search for multiferroic materials (ferroics) \cite{Schmid_2008, Saxena_2011}. Different ferroics are classified in terms of  spatial inversion and time reversal symmetry of their order parameter \cite{Spaldin_JPhys_2008}. In general, four primary orders are distinguished in ferroics: ferroelasticity, ferroelectricity, ferromagnetism, and ferrotoroidicity \cite{Loidl_JPhysCondensMatter_2008, khomskii2009trend, Gnewuch_JSolidStateChem_2019} (in what follows, the prefix ``ferro'' for the ferrotoroidicity is omitted). The magnetoelectric coupling is a secondary ferroic effect, which is inherent to toroidic order. It is of special interest in applications \cite{Li_DaltonTrans_2019}. The occurrence of multiple ferroic properties in one phase is related to specific symmetry conditions a material has to accomplish.

The lack of natural ferroic materials usable in optics has motivated a search for structures and systems that may exhibit magnetoelectric coupling arising from the metamaterial design. In particular, specifically designed particles (meta-atoms or meta-molecules) can acquire coupled electric and magnetic polarizabilities associated with the conduction currents that occur when metamaterials are irradiated by electromagnetic fields \cite{Monticone_JMaterChemC_2014}. Such particles typically are metallic split-ring resonators (SRRs) whose \textit{dynamic} magneto-optical response somewhat resembles the \textit{static} one of natural ferroic substances \cite{Pendry_1999, Zhou_PhysRevLett_2005, Wegener_PhysRevB_2009, Zheludev_Science_2010, Savinov_PhysRevB_2014, Singh_AdvMat_2018}. To describe the properties of such metamaterials composed of SRRs, the group-theoretical approach can be used. It allows one to calculate the electromagnetic modes of a resonator and determine whether this resonator exhibits a desired magneto-optical response \cite{Wongkasem_2006, Baena_ApplPhysLett_2006, Baena_PhysRevB_2007, Wongkasem_2006, Padilla_OptExpress_2007, dmitriev2009symmetry, Dmitriev_Metamat_2011, Dmitriev_IEEEAntennas_2013, Reinke_PhysRevE_2011}. Nevertheless, as far as we know, such an approach has not previously been used to describe the properties of SRR-based metamaterials exhibiting toroidicity.

Recent progress in the area of metamaterials is related to the investigation of their all-dielectric implementations \cite{evlyukhin2010optical,jahani_NatNano_2016, kruk_acsphotonics_2017,babicheva_l_appl_phys2021} that are promising candidates to overcome some issues inherent to metallic SRR-based metamaterials at higher frequencies. In  all-dielectric metamaterials,  Mie resonances of dielectric nanoparticles caused by displacement currents provide an alternative route for achieving dynamic magneto-optical response \cite{evlyukhin2012demonstration,Miroshnichenko_ACSNano_2012,Miroshnichenko_NewJPhys_2017,Lepeshov_ACSPhotonics_2018,terekhov2019multipole}. Symmetry of the dielectric resonators is tightly related to the mode structure which determines the electromagnetic features of the resonator. The structure of fields can be derived using the  group-theoretical approach \cite{Sadrieva_PhysRevB_2019, Gladyshev_PhysRevB_2020}.

Group-theoretical methods are especially useful for metamaterials whose constitutive particles have a complex shape \cite{Xiong_OptExpress_2020} or possess an in-plane broken symmetry \cite{Khardikov_JOpt_2012, Tuz_OptExpress_2018, Sayanskiy_PhysRevB_2019, Kupriiannov_OptLett_2020, Evlyukhin_PhysRevB_2020}. The resonant conditions in such metamaterials are stipulated by inherently nonradiating symmetry protected (dark) modes existing in the particles. Such modes become weakly radiative when the symmetry of the particles is broken (such symmetry protected states are also referred to as trapped modes \cite{Fedotov_PhysRevLett_2007} or, recently, as bound states in the continuum (BICs) \cite{Koshelev_PhysRevLett_2018}). The coupling to these trapped modes with incoming radiation can be controlled by the strength of asymmetry introduced in the particle. After a group-theoretical analysis of asymmetric particles, it is revealed that the dynamic magneto-optical response of the entire metamaterial can resemble the characteristics of ferroics with the magnetic dipole arrangement in either ferromagnetic or antiferromagnetic order when the trapped mode is excited \cite{Dmitriev_JApplPhys_2019, Tuz_PhysRevApplied_2020}.

The mechanism of symmetry reduction for the trapped mode excitation covered by the group-theoretical approach is also applicable to all-dielectric metamaterials composed of particle clusters \cite{Overvig_PhysRevB_2020}. In the context of ferroics, such metamaterials can possess toroidicity arising from the collective modes (supermodes) of the clusters \cite{Zheludev_PhysRevX_2015, Tasolamprou_PhysRevB_2016, Tasolamprou_chapter_2020, Xu_AdvOptMater_2019, Zhang_AdvTheorySimul_2019}. Due to low inherent losses in constituent materials, the symmetry-protected toroidal modes in all-dielectric metamaterials demonstrate a very high quality factor (high-$Q$) resonant response accompanied by the near-surface confinement of the strong electromagnetic field \cite{Tuz_ACSPhotonics_2018,  Kupriianov_PhysRevApplied_2019, Tuz_ACSApplNano_2020}. Remarkably, the conditions for the appearance of toroidal dipole modes in cluster-based all-dielectric metamaterials can be expressed in the explicit electromagnetic formulation and described in the group-theoretical language \cite{Dmitriev_2020}.

In the present paper, we aim to introduce a special approach based on the magnetic group theory to describe characteristics of the toroidal supermodes, which are implemented in cluster-based all-dielectric metamaterials. The constitutive cluster of these metamaterials has a hexagonal geometry, which is a key design in the topological photonics \cite{Wu_PhysRevLett_2015, Gorlach_2018, Yang_PhysRevLett_2018, Chern_2019, Xiong_2019, Xi_2020}. We demonstrate that in the given metamaterials, it is possible to realize excitation of the toroidal supermodes possessing either symmetric order (toroidal order, TO), when the toroidal dipole moments of all trimers are parallel, or antisymmetric order (antitoroidal order, ATO), when the toroidal dipole moments in the neighboring trimers are antiparallel having a staggered distribution. We define the selection rules and show in detail how these resonances can be excited by a specific breaking of the metamaterial unit cell symmetry. We perform a set of numerical simulations supported by microwave experiments to reveal specific physical properties of such metamaterials. 

The rest of the paper is organized as follows. Section~\ref{sec:multipoles} introduces the concept of a toroidal dipole moment existing in a trimer of dielectric disks. Then, in Sec. \ref{sec:problem}, a metamaterial composed of hexagonal unit supercells of trimers is designed to obtain a system supporting a net toroidal dipole moment. In Sec. \ref{sec:supercell} the symmetry analysis of the hexagonal unit supercell is performed. We explain what perturbations should be introduced into the supercell to excite the  toroidal dipole moments of the hexagonal cluster by the field of a normally incident linearly polarized wave. The theory of magnetic groups applied to the hexagonal structure is presented in Sec. \ref{sec:magngroup}. This theory predicts that the  toroidal dipole moments of the supercell can appear either in symmetric or antisymmetric order. The selection rules for these orders are derived in Sec. \ref{sec:rules}. Theoretical description and numerical simulations of excitation of the symmetric and antisymmetric orders in arrays due to specific perturbations of their hexagonal unit supercell are given in Secs. \ref{sec:descriptions} and \ref{sec:simulations}, respectively. In Sec. \ref{sec:experiments}, the verification of theory with a microwave experiment is provided. The presented results are discussed and  summarized in Secs. \ref{sec:discussions} and \ref{sec:conclusions}, respectively.

\section{Toroidal dipole mode of trimer}
\label{sec:multipoles}

Clusters of dielectric particles (meta-molecules) possess a set of natural modes (eigenmodes) that are defined by the cluster symmetry and obey group theory rules. This is  similar to the case of conventional molecules. The eigenmodes can be related to coefficients of the multipole decomposition of the corresponding Mie solution derived for the problem of particles interacting with an electromagnetic field \cite{Bohren_book_1998}. When the particles are arranged into a cluster, the eigenmodes of individual particles strongly interact and form supermodes of the cluster. A particular supermode can be designated as bonding or antibonding, depending on whether the induced coupling between the particles in the cluster appears in the low- or high-energy configuration \cite{Rechberger_OptCommun_2003, Chuntonov_NanoLett_2011}. The exact arrangement of the dielectric particles within a cluster therefore has an essential impact on the symmetry of the resulting supermode.

In the present study, we consider a complex hexagonal cluster consisting of six trimers of dielectric disk-shaped particles. Thus, we have the following interaction hierarchy: the eigenmodes of three individual disks are coupled into the trimer eigenmodes, which are then coupled into the hexagonal cluster supermode. We are interested in a particular eigenmode of the trimer, which exhibits a significant contribution from the toroidal dipole moment \cite{Xu_AdvOptMater_2019, Tuz_ACSApplNano_2020, Dmitriev_2020}. Then we consider the coupling of these toroidal eigenmodes within the hexagonal cluster which can be realized as a bonding (antitoroidal) or antibonding (toroidal) state.

One should note that the analysis of eigenmodes of a complex system of dielectric particles does not necessarily require the involvement of the concept of a toroidal moment. For instance, a mode coupling theory used for designing dielectric antennas and microwave circuits can be considered as an alternative approach \cite{Wang_IEEE_2007, Trubin2015lattices, Fesenko_OptExpress_2019}. Nevertheless, we are convinced that our description based on the toroidal modes makes it possible to obtain a clear physical picture of the realization of a specific resonant state in the system under consideration.

In what follows we analyze the conditions of appearance of a toroidal dipole moment in a single trimer of dielectric disk-shaped particles. Each disk in the trimer is situated in the vertex of an equilateral triangle, therefore the trimer symmetry is described by the point group $C_{3v}$. The triangle side size is $a_d$. The radius and thickness of the disks are $r_d$ and $h_d$, respectively. The disks are made of a nonmagnetic material with permittivity $\varepsilon_d$ and arranged in a homogeneous ambient space with permittivity $\varepsilon_s=1$.

A toroidal dipole moment of a system with electric current density distribution ${\bf j}({\bf r},\omega)$ (where $\bf r$ is the radius-vector of a volume element and $\omega$ is the angular frequency) is governed by the following equation \cite{Dubovik_PhysRep_1990, marinov2007toroidal, chen2011optical, evlyukhin2016optical}
\begin{equation}
    \label{TD}
{\bf T}=\frac{1}{10}\int_V [({\bf r}\cdot{\bf j}){\bf r}-2r^2{\bf j}]d{\bf r},
\end{equation}
where $V$ is the volume occupied by the current density. Here we omitted the arguments of the current density  ${\bf j}({\bf r},\omega)$ and consider that the toroidal dipole moment is located at the origin of the chosen coordinate frame. Equation (\ref{TD}) can be directly obtained from the Cartesian multipole decomposition of the current density where toroidal dipole moment presents a third-order multipole term \cite{chen2011optical, evlyukhin2016optical}. This multipole decomposition is based on the Taylor series of the Dirac $\delta$-function.

\begin{figure}[t]
\centering
\includegraphics[width=\linewidth]{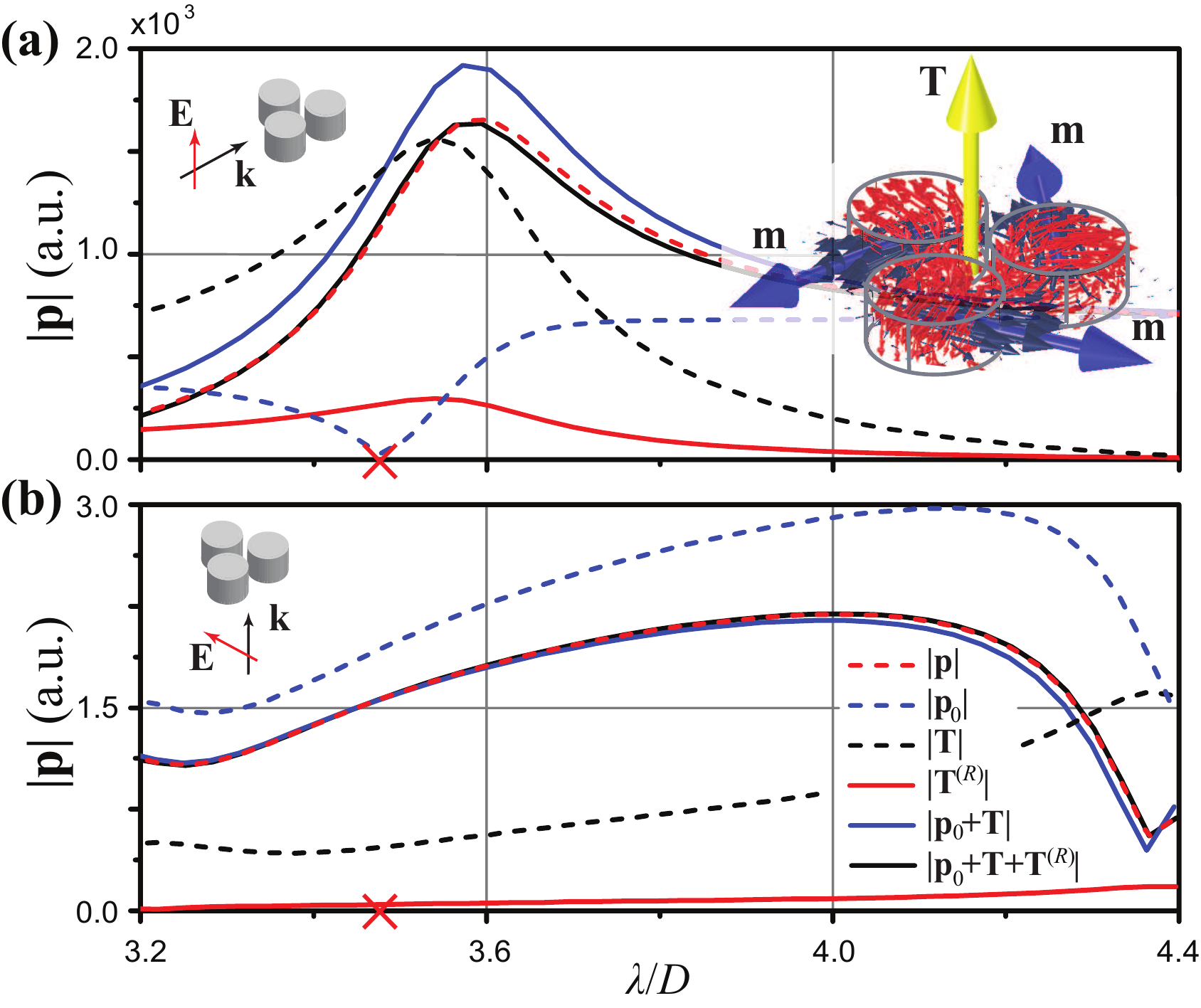}
\caption{Exact electric dipole moment with corresponding contributions of the LWA electric dipole and toroidal dipole moments of a single trimer for its (a) lateral and (b) frontal irradiation by a linearly polarized wave. The designation of the vectors $\bf k$ and $\bf E$ of the incident wave and an appearance of the trimer toroidal eigenmode are given in the inserts. In the eigenmode, red and black arrows correspond to the flow of electric polarization currents and magnetic field respectively, the bold yellow arrow indicates the toroidal dipole moment $\bf T$, and the bold blue arrows indicate the magnetic dipole moments $\bf m$ of individual disks. The eigenmode position on the wavelength scale is marked by a red cross. Parameters of the trimer are: $\varepsilon_d=22$, $h_d/D=0.45$, and $a_t/D=1.125$.}
\label{fig:fig_1}
\end{figure}

In the approach based on the spherical harmonic expansion of radiated (scattered) waves \cite{Alaee_OptCommun_2018, evlyukhin2019multipole}, the explicit contribution of the toroidal dipole moment in the multipole decomposition does not appear, and it is associated with the exact electric dipole moment \cite{fernandez2017dynamic}: 
\begin{equation}
    \label{EED}
    {\bf p}=\frac{i}{\omega}\int_V\left(j_0(kr){\bf j}+\frac{k^ 2}{\omega}\frac{j_2(kr)}{(kr)^2}[3({\bf r}\cdot{\bf j}){\bf r}-r^2{\bf j}]\right)d{\bf r},
\end{equation}
where $j_0(kr)$ and $j_2(kr)$ are the spherical Bessel functions of the zero- and second-order, respectively, $k$ is the wave number in surrounding  medium. Applying the corresponding Taylor expansions to the Bessel functions in Eq.~(\ref{EED}) and writing out the first three terms  explicitly, one obtains  \begin{eqnarray}\label{EED1}
    {\bf p}&=&\frac{i}{\omega}\int_V{\bf j}d{\bf r}+\frac{ik}{c}\frac{1}{
    10}\int_V[({\bf r}\cdot{\bf j}){\bf r}-2r^2{\bf j}]d{\bf r}\nonumber\\
    &&+\frac{ik^3}{c}\frac{1}{280}\int_V [3r^4{\bf j}-2r^2({\bf r}\cdot{\bf j}){\bf r}]d{\bf r}+\ldots\nonumber\\
    &=&{\bf p}_0+ \frac{ik}{c} {\bf T}+ \frac{ik^3}{c}{\bf T}^{(R)}+\ldots
\end{eqnarray}
where $c$ is the light speed in surrounding medium, ${\bf p}_0$ is the Cartesian electric dipole moment obtained from the $\delta$-function Taylor expansion corresponding to the long-wavelength approximation (LWA)  \cite{Alaee_OptCommun_2018}, ${\bf T}^{(R)}$ is the mean-square radius of the toroidal moment \cite{nemkov2018electromagnetic,Gurvitz_LPOR_2019}. 

From Eq.~(\ref{EED1}) one can consider the toroidal dipole moment as only a next correcting term to the Cartesian electric dipole moment ${\bf p}_0$ used for calculation of the exact electric dipole moment $\bf p$, and, thus, the toroidal dipole moment cannot be considered separately. However, this statement does not reflect total physical role of the toroidal dipole moment. Indeed, depending on the frequency of electromagnetic fields, shape, size, and material parameters of an electric current system, it can appear that the first term in Eq.~(\ref{EED1}) is very small, or even equals to zero, so that the main contribution goes from the toroidal term ($ik{\bf T}/c$). In this case, the electric current systems can be considered as supporting a purely electric toroidal response, and, thus, the toroidal dipole moment acquires an independent physical meaning.

To demonstrate this peculiarity, we consider the trimer irradiation by an electromagnetic plane wave from both lateral and frontal directions. The absolute value of the exact electric dipole moment $\bf p$ and the corresponding contributions of the LWA electric dipole ${\bf p}_0$ and the toroidal dipole moments $\bf T$ together with the mean-square radius ${\bf T}^{(R)}$ are presented in Fig. \ref{fig:fig_1} for two irradiation conditions of the linearly polarized plane wave. Calculation procedure corresponds to the description from \cite{Tuz_ACSApplNano_2020}.  In our calculations, we normalize the incident wavelength and all geometrical parameters of the problem on the disk diameter $D=2r_d$.

One can see that at the lateral irradiation condition [Fig. \ref{fig:fig_1}(a)], the exact dipole moment of the trimer is resonantly excited. Decomposition of its value on the basis of Eq. (\ref{EED1}) shows that the toroidal dipole $\bf T$ provides the main contribution in the resonant exact electric dipole moment $\bf p$ [the corresponding wavelength is indicated by a red cross in Fig. \ref{fig:fig_1}]. In this case, the electric dipole response of the trimer is associated with excitation of the trimer toroidal dipole mode [see the inset with the trimer toroidal eigenmode in  Fig. \ref{fig:fig_1}(a)]. For the other irradiation conditions [Fig. \ref{fig:fig_1}(b)] the toroidal dipole mode is not excited. In this case, the exact dipole moment $\bf p$ is basically determined by the interference between the ${\bf p}_0$ and $\bf T$ dipole moments. Note that in this condition the significant suppression of $\bf p$ appears at the region of large wavelengths as a result of destructive interference. This state can be attributed to  the anapole state \cite{zenin2017direct, yang2019nonradiating, baryshnikova2019optical, savinov2019optical}.

Therefore, it is revealed that for the given eigenmode of the trimer, the toroidal dipole moment provides a dominant contribution, and this eigenmode can arise only for a certain type of excitation of the incident field. In  particular, this toroidal mode can be considered as a dark and bright state for the field of frontally and laterally incident wave, respectively. In what follows, we show the mechanism of access to this indicated dark state under frontal irradiation conditions in metamaterials composed of hexagonal clusters of trimers by breaking their symmetry.
 
\section{Metamaterial with hexagonal unit supercell}
\label{sec:problem}

\begin{figure}[t]
\centering
\includegraphics[width=\linewidth]{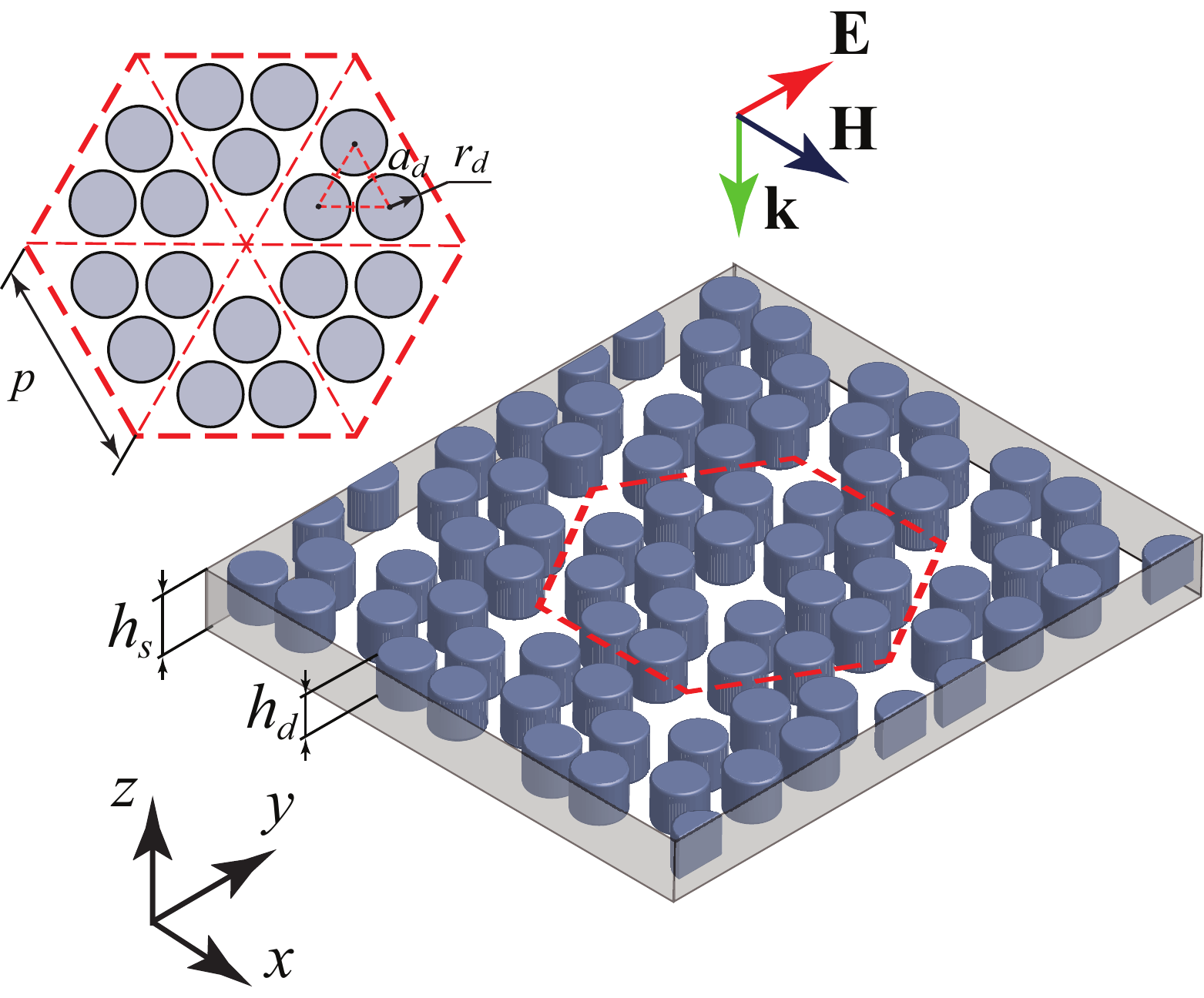}
\caption{Coordinate frame and schematic view of a dielectric hexagonal array composed of trimer-based supercells. The unit supercell is outlined by a red dashed contour.}
\label{fig:fig_2}
\end{figure}

We consider a metamaterial composed of a planar array with trimers of dielectric disks (i.e., the trimer is a minicell of the metamaterial). The six trimers are arranged in such a way that hexagonal clusters appear (i.e., the hexagonal cluster is a supercell of the metamaterial). The hexagon lateral size is $p$. The array of disks is situated in the $x$-$y$ plane as it is shown schematically in Fig.~\ref{fig:fig_2}. To form a metamaterial, the dielectric disks are immersed symmetrically (preserving the plane of symmetry $z=0$) into a dielectric substrate with relative permittivity $\varepsilon_s$ and thickness~$h_s$.

A peculiarity of this new type of metamaterial is a highly symmetric hexagonal geometry of its unit supercells based on dielectric trimers. The optical response of the entire metamaterial is largely determined by the symmetry of the supercell, which can be deliberately lowered by introducing certain perturbations into the hexagonal cluster.

Our goal is to find specific perturbations so that the toroidal dipole mode of the trimers can be effectively excited in the metamaterial by the field of a normally incident linearly polarized wave with the wave vector ${\bf k} = \{0,0,-k_z\}$ and wavelength $\lambda=2\pi c/\omega$. We define that the metamaterial is irradiated by either $x$-polarized (${\bf E},{\bf H} = \{E_x,H_y$,0\}) or $y$-polarized (${\bf E},{\bf H} = \{H_x,E_y,0\}$) wave. Due to  nature of the toroidal dipole mode, it is appropriate to consider the problem using the magnetic field ${\bf H}$ of the incident wave as the exciting source. 

\section{Supercell symmetry analysis} 
\label{sec:supercell}
For the given planar arrays of dielectric disks, one can consider the problem in the framework of two-dimensional (2D) symmetry. Since the center of the equilateral triangle of the underlying trimer and the center of the minicell are the same, the minicell geometry corresponds to the $C_{3v}$ symmetry. To get access to the toroidal dipole mode, the minicell symmetry must be reduced \cite{Dmitriev_2020}. This symmetry reduction can be realized in several ways. In particular, the thickness of particular disks in the trimers can be resized (out-of-plane perturbation) or these disks can be shifted aside (in-plane perturbation). 

The geometry of the hexagonal unit supercell corresponds to the $C_{6v}$ symmetry. As a main reference point, the scheme of the supercell with planes of symmetry and rotational elements of the $C_{6v}$ group are presented in Fig.~\ref{fig:fig_3} (the Sch\"oenflies notation of elements of point symmetry is given in Appendix \ref{app:Schoenflies}). 
\begin{figure}[t]
\centering
\includegraphics[width=60mm]{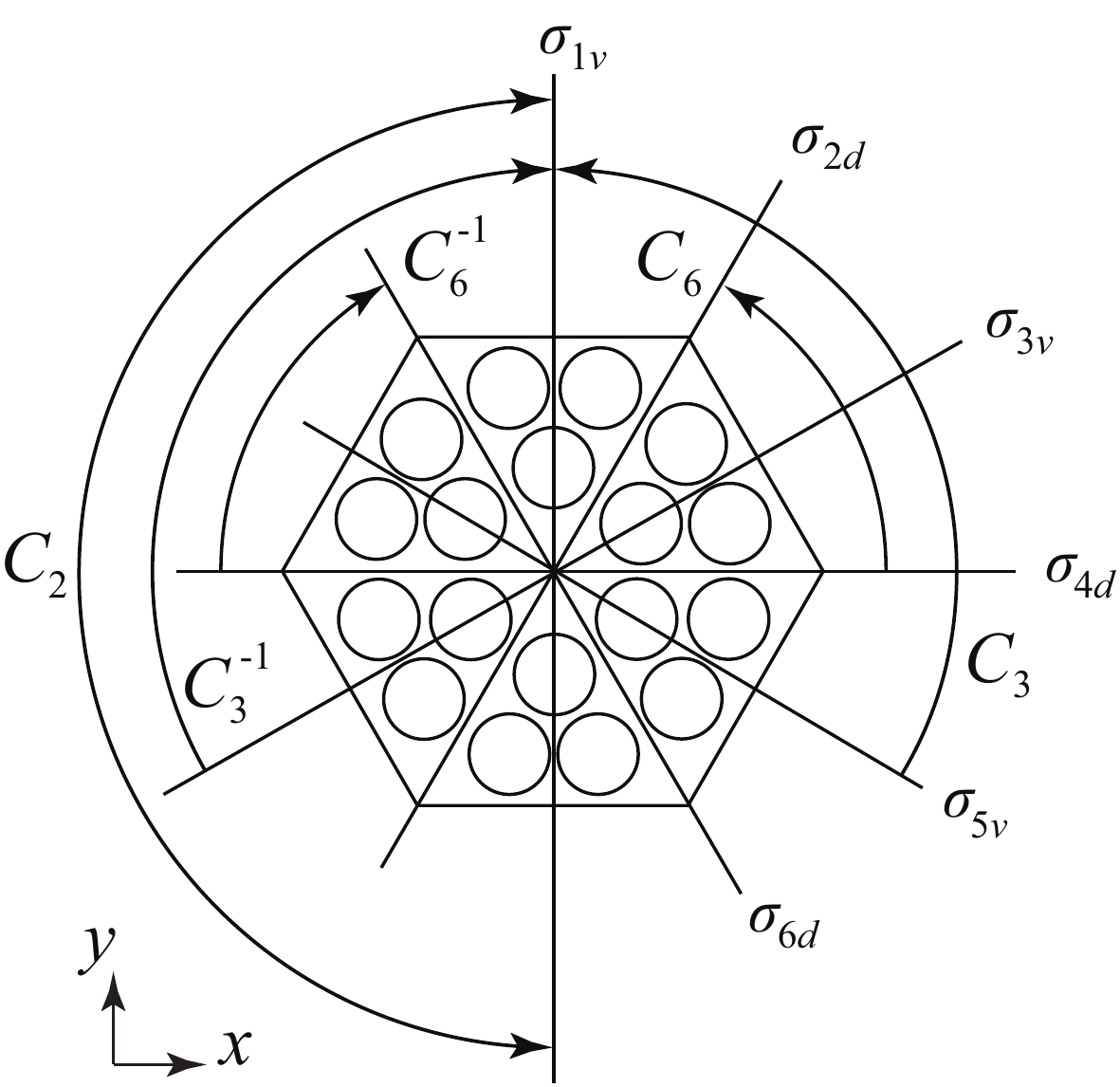}
\caption{Hexagonal supercell described by the $C_{6v}$ group and elements of this group.}
\label{fig:fig_3}
\end{figure}
\begin{figure*}[t]
\centering
\includegraphics[width=165mm]{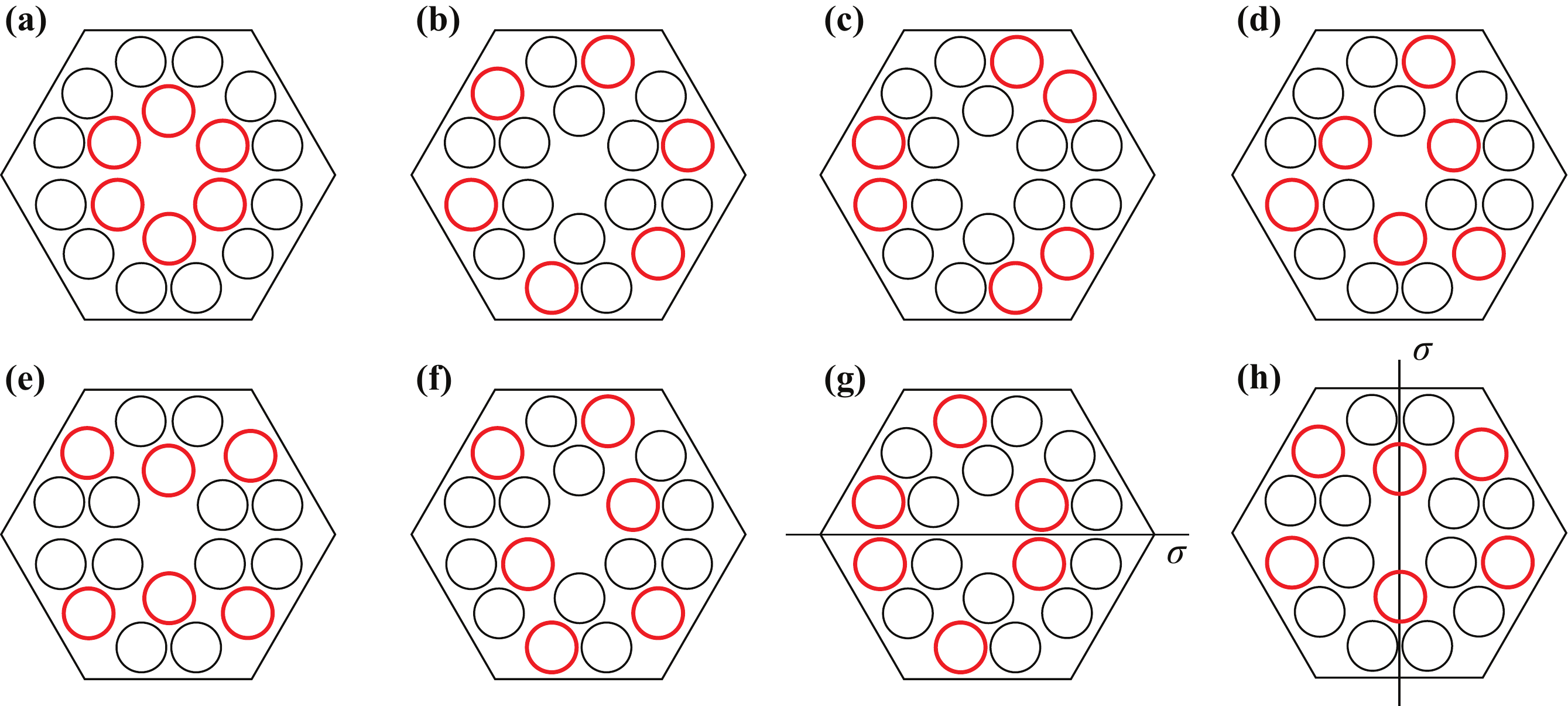}
\caption{Examples of the perturbed hexagonal unit supercells whose symmetry is reduced to the groups: (a)~$C_{6v}$, (b)~$C_{6}$, (c)~$C_{3v}$, (d)~$C_{3}$, (e)~$C_{2v}$, (f)~$C_{2}$, (g)~$C_{s}^{(1)}$, and (h)~$C_{s}^{(2)}$. Perturbed disks are denoted by red circles.}
\label{fig:fig_4}
\end{figure*}
With such a  supercell symmetry, the toroidal dipole mode is a dark state of the hexagonal cluster and cannot be excited by the field of a normally incident linearly polarized wave. Therefore, the symmetry $C_{6v}$ must be reduced. Particular perturbations lead to different manifestations of the toroidal dipole mode in the metamaterial, which can arise in either symmetric or antisymmetric order.  The relationship between the perturbed supercell symmetry and the toroidal dipole mode order is the subject of our subsequent study. 

Here we introduce a classification of possible reductions of the  supercell symmetry, assuming that only one disk in each trimer forming the hexagonal cluster can be perturbed and only one type of perturbation is allowed. 
With a set of such allowed perturbations, one can design supercells having the highest $C_{6v}$ symmetry as well as the symmetries of the $C_{6v}$ subgroups (the subgroup decomposition of the $C_{6v}$ group with the $C_{6}$, $C_{3v}$, $C_{3}$, $C_{2v}$, $C_{2}$, and $C_{s}$ subgroups is illustrated in Fig.~\ref{fig:fig_A1} of Appendix \ref{app:Schoenflies}). 

The unique way to obtain the $C_{6v}$ symmetry in the perturbed supercell is shown in Fig. \ref{fig:fig_4}(a). In this geometry, all six perturbed disks are closest to the center of the hexagonal supercell. The perturbed supercells with the $C_{6}$, $C_{3v}$, and $C_{3}$ symmetries are shown in Figs.~\ref{fig:fig_4}(b)-\ref{fig:fig_4}(d). Notice that these geometries are not unique and the positions of the perturbed disks can be different. The geometries of the supercells with the $C_{2v}$ and $C_{2}$ symmetries are presented in Figs.~\ref{fig:fig_4}(e) and \ref{fig:fig_4}(f), respectively. These geometries are also not unique.

The $C_{s}$ symmetry can be realized in two variants, where the plane of symmetry $\sigma$ passes through an apex [Fig. \ref{fig:fig_4}(g)] or between two apexes [Fig. \ref{fig:fig_4}(h)] of the hexagon. These variants are designated as the $C^{(1)}_{s}$ and $C^{(2)}_{s}$ symmetries, respectively. Again, different geometries are possible in the hexagonal clusters with the $C^{(1)}_{s}$ and $C^{(2)}_{s}$ symmetries.

\section{Magnetic group description} 
\label{sec:magngroup}
%

We investigate  electromagnetic response of the given metamaterial to external excitation by an alternating electromagnetic field. However, at a certain point in time, one can get a fixed picture of the field distribution in the unit supercell. This field pattern is associated with the characteristics of both the  eigenmode that is excited and the magnetic field vector of the incident wave. Thus one can construct a general theory which combines the geometric symmetry of the unit supercell and the ``dynamic'' symmetry of the alternating magnetic fields and toroidal moment (see Appendix D).

\begin{figure}
\centering
\includegraphics[width=55mm]{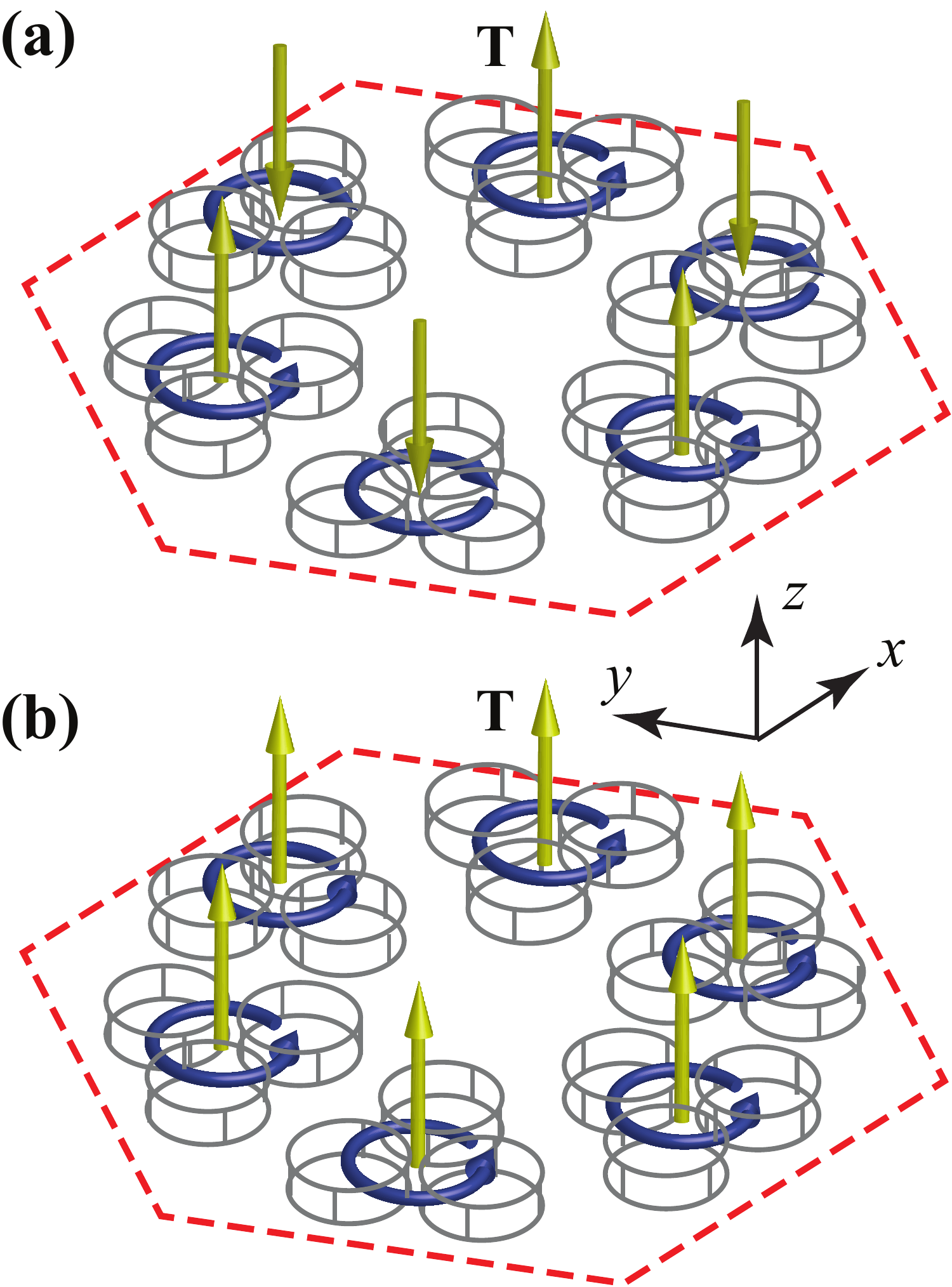}
\caption{Schematics of (a) ATO and (b) TO modes in a hexagonal unit supercell. Blue and yellow arrows demonstrate the magnetic field flow and orientation of the toroidal dipole moments, respectively.}
\label{fig:fig_5}
\end{figure}

In the  eigenmode of  unperturbed trimer, the toroidal dipole moment $\bf T$ has only single component $T_z$ \cite{Dmitriev_2020}. In a fixed moment of time, the normalized vector $\bf T$ can acquire only two states being oriented either ``up'' ($T_z=+1$) or ``down'' ($T_z=-1$). Accordingly, in a set of two trimers, two combinations of the toroidal dipole moments are possible when they are paired in either bonding (contra-directional) or antibonding (co-directional) fashion. When considering the hexagonal unit supercell,  we are interested in two specific toroidal eigenmodes, whose  moments are oriented either in the staggered order or in the same direction in all six trimers (see Fig.~\ref{fig:fig_5}). In what follows, we distinguish these eigenmodes as the ATO and TO modes, respectively.

In the symmetry analysis, these eigenmodes can be attributed to the one-dimensional (1D) irreducible representations (IRREPs) of the $C_{6v}$ group. The TO mode belongs to IRREP $A_1$, i.e., it is not changed under all the symmetry elements. The ATO mode is transformed in accordance with IRREP $B_1$ and the toroidal dipole moment changes its sign under the $C_2$, $C_6^1$, and $C_6^{-1}$ transformations, and three reflections $\sigma_d$ (see Table~\ref{tab:CharC6v} of Appendix~\ref{app:tables}).  

Notice that the in-plane magnetic dipole moments induced in dielectric disks by the external magnetic field can be used to construct an alternative theory. In this case, one has to work with  2D representations of the {\it axial} vector  $\bf m$ rather than with 1D IRREPs of the {\it polar} vector ${\bf T}$. However, this alternative description involves more complex transformation properties of $\bf m$ and this  complicates the analysis.

Taking advantage of the time reversal symmetry of the toroidal dipole moment $\bf T$, one can use this quantity to define the magnetic symmetry of the alternating fields in the array. Magnetic groups based on the $C_{6v}$ group of symmetry, its subgroups and elements of these groups are summarized in Table~\ref{tab:C6v}. In this table, ${\cal T}$ is the time reversal operator. 
%
\begin{table*}[htb]
\centering
\caption{Possible symmetries of toroidal dipole modes: magnetic groups based on the $ C_{6v}$ symmetry, their subgroups and elements}
\begin{tabular}{c@{\hspace{1mm}}c@{\hspace{4mm}}c@{\hspace{2mm}}c@{\hspace{2mm}}c@{\hspace{1mm}}c@{\hspace{1mm}}c@{\hspace{2mm}}c@{}}
\hline
\hline
1st category  & $C_{6v}+{\cal T} C_{6v}$ & $C_{6}+{\cal T}C_{6}$ & $C_{3v}+{\cal T}C_{3v}$ & $C_{3}+{\cal T}C_{3}$ & $C_{2v}+{\cal T}C_{2v}$ & $C_{2}+{\cal T}C_{2}$ & $C_{s}+{\cal T}C_{s}$ \\
\\
\hline
\hline
2nd category & $\bf C_{6v}$ &  $\bf C_{6}$  &  $\bf C_{3v}$ & $\bf C_{3}$ &  $\bf C_{2v}$ & $\bf  C_{2}$ & $\bf  C_{s}$\\
\hline
Content & $e$,  $C_6$, $C_6^{-1}$,  $C_2$ & $e$, $C_6$, $C_6^{-1}$ &  $e$, $C_3$, $C_3^{-1}$ &$e$, $C_3$, $C_3^{-1}$ &$e$, $C_2$ &e, $C_2$ &$e$, $\sigma$\\
& $C_3$, $C_3^{-1}$,  $3\sigma_v$, $3\sigma_d$  &  $C_3$, $C_3^{-1}$,   $C_2$ &  $\sigma_1$, $\sigma_2$, $\sigma_3$ & &   $\sigma_1$, $\sigma_2$, &  & \\
\\
\hline
\hline
3rd category &\,\,\,$C_{6v}(C_{6})$\,\,\,\,\vline \,\,   $C_{6v}(C_{3v})$  & $C_{6}(C_{3})$  & $C_{3v}(C_{3})$ & &   $C_{2v}(C_{2})$ \,\,\vline\,\, $C_{2v}(C_{s})$ & $C_{2}(C_{1})$ & $C_{s}(C_{1})$\\
\hline
Content  &\,\,\,$e$,  $C_6$, $C_6^{-1}$\,\,\vline\,\, $e$, $C_3$, $C_3^{-1}$ & $e$, $C_3$, $C_3^{-1}$  &  \, $e$, $C_3$, $C_3^{-1}$ & & $e$,  $C_2$ \,\,\vline\,\, $e$,  $\sigma_1$ & $e$, ${\cal T}C_2$ & $e$, ${\cal T}\sigma$\\
&\!\!\!$C_2$, $C_3$, $C_3^{-1}$\vline\,\,$\sigma_1$, $\sigma_2$, $\sigma_3$  &  ${\cal T}C_2$, ${\cal T}C_6$,  &   ${\cal T}\sigma_1$, ${\cal T}\sigma_2$,  &  &  ${\cal T}\sigma_1$, ${\cal T}\sigma_2$ \,\,\,\,\vline\,\, ${\cal T}C_2$, ${\cal T}\sigma_2$  &  & 
\\
& \,\,\,\,\,\,\,\,\,\,\,\, $3{\cal T}\sigma_v$  \vline\,${\cal T}C_2$, ${\cal T}C_6$  &  ${\cal T}C_6^{-1}$ & ${\cal T}\sigma_3$&&&&\\
& \,\, $3{\cal T}\sigma_d$  \,\,\vline\,\, ${\cal T}C_6^{-1}$  &   & &&&&\\
\hline
\hline
\end{tabular}
  \label{tab:C6v}
\end{table*}
In terms of magnetic groups, the  vector  $\bf T$ is odd in time. Application of the time reversal operator  ${\cal T}$ to  the toroidal dipole moment $T_z$ results in the vector $\bf T$ reorientation, ${\cal T}T_z = {-T_z}$.  

Table \ref{tab:CharC6v} of Appendix \ref{app:tables} in the first seven columns presents  space transformation properties of the polar vector $\bf T$. This transformation is defined by the formula ${T_z}^{\prime} = \chi T_z$, where $T_z$ is a given toroidal moment, ${T_z}^{\prime}$ is the mapped moment, and $\chi$ is 1D IRREP of the rotation-reflection symmetry element $R$. The eighth column of this table consists of the space-time transformation of $\bf T$ (which is an odd in time quantity) in terms of magnetic groups. Thus, Table \ref{tab:CharC6v} provides a description of two sides (two physical properties) of the toroidal moment $\bf T$. 
 
It should be emphasized that we apply the theory of magnetic groups only to the fields and toroidal dipole moments $\bf T$ of the corresponding eigenmodes. However, the possible magnetic groups are defined by  {\it geometrical} symmetry of the unit supercell.

\section{Selection rules and polarization properties of arrays}
\label{sec:rules}

A detailed description of selection rules for the quasi-bound states in the case of photonic crystal slabs is presented in \cite{Overvig_PhysRevB_2020}, where the authors intensively used the theory of {\it nonmagnmetic} groups. One of  our   motivations in this work is to show that  the  {\it magnetic} group theory gives some advantages in description of toroidal dipole mode resonances. 

The problem of the toroidal dipole mode excitation in the array of hexagonal supercells can be divided into two subproblems: (i) excitation of the toroidal dipole mode in an individual trimer and (ii) excitation of the whole hexagonal unit supercell. 

To resolve the first subproblem, one should reduce the $C_{3v}$ symmetry of the trimer to the $C_s$ group. This mechanism is studied both theoretically and experimentally in Refs.~\cite{Xu_AdvOptMater_2019, Tuz_ACSApplNano_2020, Dmitriev_2020}, so we omit the details here (for a brief description, see Appendix \ref{app:dynamic}). In the present consideration, we concentrate on solving the second subproblem, i.e., defining the selection rules for symmetries of the hexagonal supercell which allow or forbid excitation of toroidal dipole modes in the metamaterial by the field of a normally incident linearly polarized wave. 

\subsection{Selection rules for  ATO mode} 
The magnetic field $\bf H$ of the incident wave  contains the elements $e$, ${\cal T}C_{2}$, ${\cal T}\sigma_2$, and $\sigma_1$ of magnetic symmetry (see Appendix \ref{app:magnetic}). The element $C_{2}$  is incompatible with  ${\cal T}C_{2}$ of the magnetic field $\bf H$  (these elements are ``orthogonal'' in the sense that their application leads to opposite orientations of the toroidal dipole moment). Therefore, if magnetic symmetry contains the $C_{2}$ element, excitation of an ATO mode in the metamaterial with such a symmetry is forbidden. It imposes a very rigid ({\it hard}) restriction, which does not depend on the orientation of the magnetic field $\bf H$. The same is true for the groups of symmetries containing the $C_{6}$ and $C_{3}$ elements,  because the $C_{6}$ element is always accompanied by the $C_{2}$ element, whereas, from the viewpoint of polarization properties, the $C_{2}$ element is ``hidden'' in the $C_{3}$ group. However, there is no restriction on the presence of the element ${\cal T}C_{2}$.  

Now we apply to the restrictions imposing by  planes $\sigma$ and antiplanes ${\cal T}\sigma$ of symmetry. The magnetic groups of ATO mode cannot contain an element $\sigma$ coinciding with  ${\cal T}\sigma$ of the magnetic field. Analogously, the magnetic groups of the ATO mode cannot contain the element ${\cal T}\sigma$  coinciding with $\sigma$ of the magnetic field. Therefore, the vector $\bf H$ of the incident wave parallel to $\sigma$, or perpendicular to ${\cal T}\sigma$ (both $\sigma$ and ${\cal T}\sigma$ are elements of the toroidal eigenmode of the supercell) cannot be coupled to this mode. However, a small deviation of orientation of $\bf H$ from the discussed planes leads to appearance of the field $\bf H$ component which can interact with the corresponding mode of the hexagonal unit supercell. Therefore, these selection rules can be called {\it soft} restrictions.

The above analysis allows us to exclude from consideration those symmetries for which  excitation of the ATO mode in the hexagonal unit supercell is forbidden. The magnetic groups which permit excitation of the ATO mode are subgroups of the group of magnetic field $\bf H$. They are $C_{2v}(C_s)$ (the group of $\bf H$ itself), $C_{2}(C_1)$, $C_{s}(C_1)$, and $\bf C_{s}$ (see Table \ref{tab:C6v}). It means that the corresponding geometrical symmetries of the unit supercells are $C_{2v}$, $C_{2}$, and $ C_{s}$ (see corresponding perturbations of the hexagonal unit supercell in Fig. \ref{fig:fig_4}).

\subsection{Selection rules for  TO mode} 
Now we apply to the selections rules of the TO mode. After application of any of the  rotations $C_{6}$, $C_{3}$, and $C_{2}$, the sign of $\bf T$ must be preserved. Therefore, rotations with ${\cal T}$ in the magnetic groups which change the sign of $\bf T$ are prohibited (they are ${\cal T}C_{2}$ and ${\cal T}C_{6}$; the ${\cal T}C_{3}$ element does not exist in the magnetic groups). For the same reason, the ${\cal T}\sigma$ elements are also not allowed. Summarizing, the antiunitary elements cannot enter in the magnetic groups of the TO mode.

On the other hand, the elements $C_2$, $C_3$, and $C_6$ are not allowed due to their incompatibility with the field $\bf H$ symmetry. Therefore, the plane $\sigma$ is the only permissible element of symmetry, and for the mode excitation, this element must be oriented perpendicular to the field $\bf H$. Thus, the toroidal dipole mode can be excited in the unit supercell with $\bf {C_s}$ symmetry of the magnetic groups, which contains the element $\sigma$. This condition is similar to that defining the excitation of the toroidal dipole mode in an isolated trimer \cite{Dmitriev_2020}.

Two schemes of perturbation of the unit supercell described by the $C_s$ group are shown in Figs.~\ref{fig:fig_4}(g) and \ref{fig:fig_4}(h).  Among them, the scheme presented in Fig.~\ref{fig:fig_4}(g) permits excitation of both the ATO and TO modes. 
\subsection{Polarization properties of arrays with ATO and TO modes} 
Polarization properties of the given arrays depend on their symmetry conditions. As it is known, for the polarization insensitivity the object must have the rotational symmetry with the axis $C_3$, $C_4,\ldots, C_\infty$ \cite{Mackay_ElectronLett_1989}. From the foregoing analysis, it follows that the symmetries allowing the excitation of the ATO and TO modes do not contain  rotational elements $C_n$ with $n>2$. Therefore, polarization insensitivity of the arrays is not possible for both the ATO and TO modes. 
\section{Specific arrays description}
\label{sec:descriptions}
\subsection{ATO mode in  $C_{2v}$ supercells} 
From the preceding consideration of the selection rules it follows that to get access to the ATO mode, symmetry of the hexagonal unit supercell needs to be reduced to the $C_{2v}(C_s)$ groups [the corresponding unit supercell perturbation is presented in Fig.~\ref{fig:fig_4}(e)]. The  $C_{2v}(C_s)$ group of the third category does not contain the isolated time reversal operator ${\cal T}$. Thus, at an arbitrary fixed moment in time, changing the sign of time inverts $T_z$.

In the group-theoretical language, the ATO mode belongs to the IRREP $B_1$ of the $C_{6v}$ group (see Table \ref{tab:CharC6v} in Appendix \ref{app:tables}). In this symmetry, the toroidal dipole mode is a dark state. The IRREP $B_1$ of the $C_{6v}$ degenerates in the IRREP $B_1$ of the lower $C_{2v}$ group (see Tables \ref{tab:ERREPC2v} and  \ref{tab:degeneration} in Appendix \ref{app:tables}). The geometrical symmetry described by the $C_{2v}$ group is consistent with the $C_{2v}(C_s)$ symmetry of the incident magnetic field $\bf H$. Therefore, in this case, one can expect an efficient excitation of the toroidal dipole mode by a normally incident electromagnetic wave with a proper polarization.  

To obtain the high-$Q$ resonant conditions, the introduced perturbation in the unit supercell should be small. When the perturbation is small, the characteristics of the local electromagnetic fields in the array with the $C_{2v}$ symmetry does not differ significally from those of the $B_1$ dark eigenmode in the non-perturbed array with the $C_{6v}$ symmetry. One can conclude, that the approximate symmetry of the electromagnetic field in the hexagonal unit supercell with the $C_{2v}$ symmetry corresponds to the IRREP $B_1$ of the $C_{6v}$ group or, in the framework of magnetic groups, to the $C_{6v}(C_{3v})$ group.

\subsection{ATO and TO modes in  $C_{s}$ supercells}

Reduction of the $C_{6v}$ symmetry to the $ C_{s}$ group allows one to excite the ATO and TO modes (see Tables \ref{tab:IRREPCs} and \ref{tab:degeneration} in Appendix \ref{app:tables}). Besides the unit element $e$, the $C_{s}$ group contains only one plane of symmetry $\sigma$. This plane can either lie between the trimers [Fig.~\ref{fig:fig_4}(g)] or pass through the centers of the trimers [Fig.~\ref{fig:fig_4}(h)]. Recall, that these geometries are not unique, some other positions of the perturbed disks can also result in the $C_{s}$ symmetry.

\begin{figure}[t!]
\centering
\includegraphics[width=\linewidth]{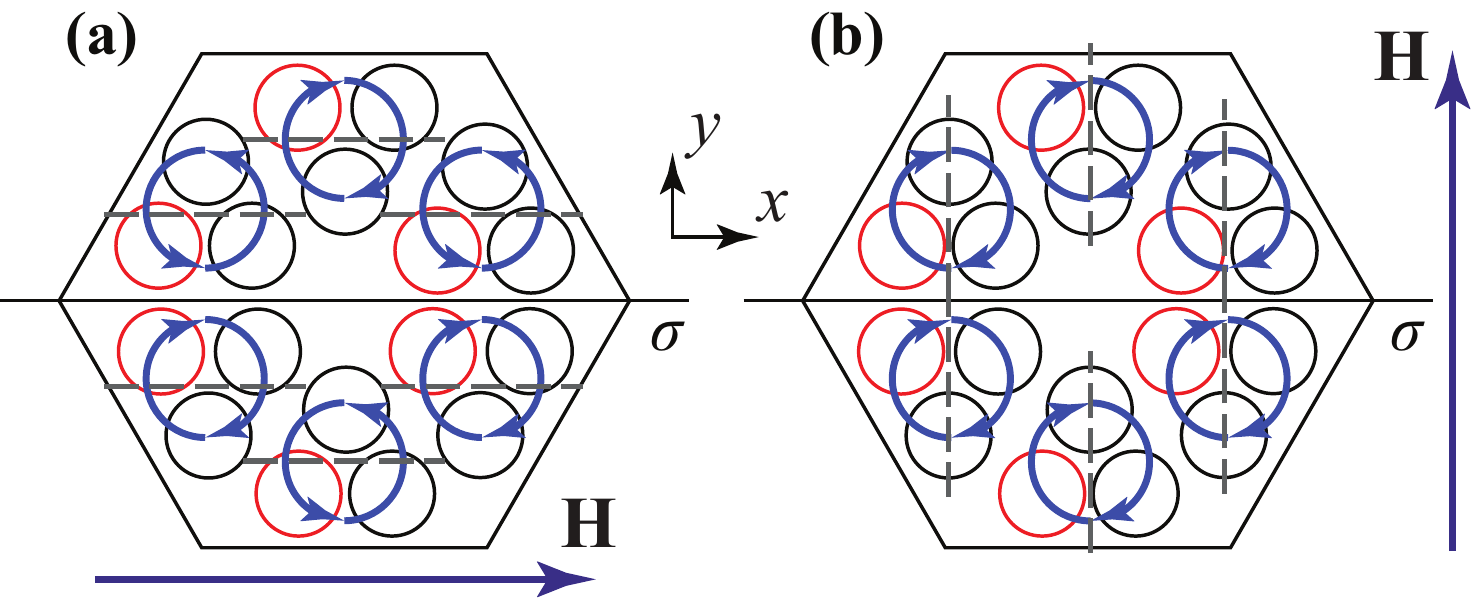}
\caption{Excitation of (a) ATO mode and (b) TO mode in the same hexagonal supercell whose symmetry is reduced to the $C_{s}^{(1)}$ group. Perturbed disks are denoted by red circles. Orientation of the magnetic fields $\bf H$ of the incident wave and flow of the inner magnetic field of the toroidal dipole eigenmode are given by the dark and light blue arrows, respectively. The test lines are denoted by grey color.}
\label{fig:fig_6}
\end{figure}

Noteworthy that if the magnetic field $\bf H$ of the incident wave is orthogonal to $\sigma$ [see Fig.~\ref{fig:fig_6}(a)], the ATO mode can be excited, whereas for the same geometry but with the magnetic field $\bf H$ parallel to $\sigma$ [see Fig.~\ref{fig:fig_6}(b)], the TO mode arises. 
\subsection{Rule of thumb for  mode order} 
To determine the direction of in-plane rotation of the magnetic field in the trimers and therefore the orientation of the out-of-plane toroidal dipole moments in the array, the following simple rule of thumb can be used. One should draw test lines through the centers of all trimers in the unit supercell parallel to the magnetic field $\bf H$. If the perturbed disk of a given trimer is on one side of the test line, it corresponds to a certain direction of rotation of the local magnetic field of the corresponding mode. If the perturbed disk is on the other side of the test line, this gives the opposite direction of rotation of the local magnetic field. This allows one to define immediately the  order of the excited net toroidal dipole mode. If the test line passes through the center of the perturbed disk, this resonator cannot be excited directly by the incident field. However, it can be excited indirectly via the neighboring resonators due to the coupling effect between particles in the array.

This intuitive picture can be understood from Fig.~\ref{fig:fig_6}, where the hexagonal unit supercell with the $C_{s}^{(1)}$ symmetry is presented. When the vector ${\bf H}$ is parallel to the $x$ axis [Fig.~\ref{fig:fig_6}(a)], the perturbed disks appear at different sides of the test lines which results in the staggered orientation of the toroidal moments in the neighboring trimers, i.e., the ATO mode can be excited. Contrariwise, for the same unit supercell but when the vector ${\bf H}$ is parallel to the $y$ axis [Fig.~\ref{fig:fig_6}(b)], the structure produces parallel orientation of the toroidal moments, i.e., the TO mode can be excited. 
\section{Numerical  results}
\label{sec:simulations}
We now proceed to check our group-theoretical description of the excitation of toroidal dipole modes in the discussed arrays by performing the full-wave numerical simulations. To calculate both eigenmodes and transmitted spectra of the array irradiated by a linearly polarized plane wave, we use the RF module of the commercial COMSOL Multiphysics finite-element electromagnetic solver. In the solver, we impose the Floquet-periodic boundary conditions on four sides of the unit cell to simulate an infinitely extended in the $x$-$y$ plane arrangement of dielectric disks. The code related to the multipole decomposition with accounting for the toroidal dipole term defined by Eq. (\ref{TD}) is incorporated into the solver as a special component. The solver allows us to plot the distribution of the inner electromagnetic field within the hexagonal unit supercell at a specified resonant wavelength.

Initially we calculate the eigenmodes that exist in a reference metamaterial composed of unperturbed supercells. The material losses in the disks are excluded in these simulations. From the released results, two specific eigenmodes are selected, which demonstrate the appearance of the parallel or antiparallel orientation  of the toroidal dipoles. Then the transmitted spectra of the metamaterial are calculated in the wavelength range where the selected eigenmodes exist. Both $x$-polarized and $y$-polarized waves are under consideration. 

\begin{figure}[t!]
\centering
\includegraphics[width=\linewidth]{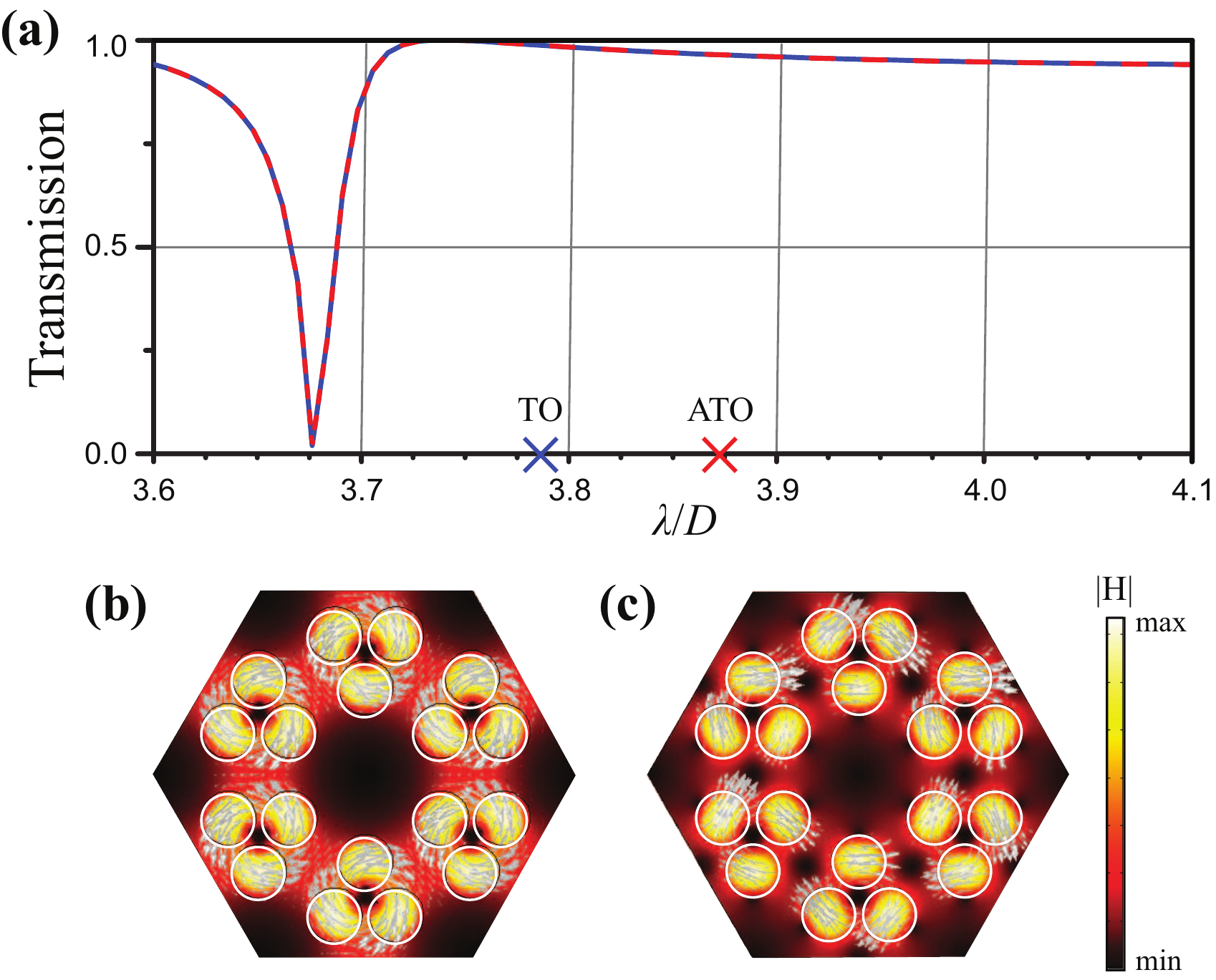}
\caption{(a) Transmitted spectra of the metamaterial with unperturbed unit supercells excited by the $x$-polarized (solid blue lines) and $y$-polarized (dashed red lines) wave. The normalized magnitude and flow of the magnetic near-field of the (b) ATO and (c) TO modes. The material and geometrical parameters of the trimers are the same as in Fig.~\ref{fig:fig_1} and $p/D=3.375$.}
\label{fig:fig_7}
\end{figure}

The results of our simulations of characteristics of the reference metamaterial are presented in Fig.~\ref{fig:fig_7}. As before, we present our results in the dimensionless parameters, where all values are related to the disk diameter $D$. The eigenwave solution shows that the corresponding mode of the trimer exists in the array in the ATO and TO states, whose resonant wavelengths are different from the resonant wavelength of the toroidal dipole mode of a single trimer. 

One can conclude, that the spectral characteristics of the metamaterial are the same for waves of both polarizations, and while the symmetry of the supercells is preserved, there are no peculiarities in the transmitted spectra at the wavelengths of the ATO and TO modes. This confirms the dark feature of these modes.

Further two particular designs of the perturbed unit supercells are chosen to show a possibility to excite the ATO and TO modes in the metamaterial. They are the supercells with the $C_{2v}$ and $C_s^{(1)}$ symmetries [see Figs. \ref{fig:fig_4}(e) and \ref{fig:fig_4}(g), respectively]. As a perturbance, we consider the change in the thickness of a corresponding disk in each trimer forming the unit supercells. This change is defined by the value $\Delta h_d$, and the dimensionless unit supercell asymmetry parameter is introduced as $\theta=(h_d+\Delta h_d)/h_d$. 

\subsection{Array with  $C_{2v}$ symmetry of unit supercell} 
\begin{figure}[t]
\centering
\includegraphics[width=\linewidth]{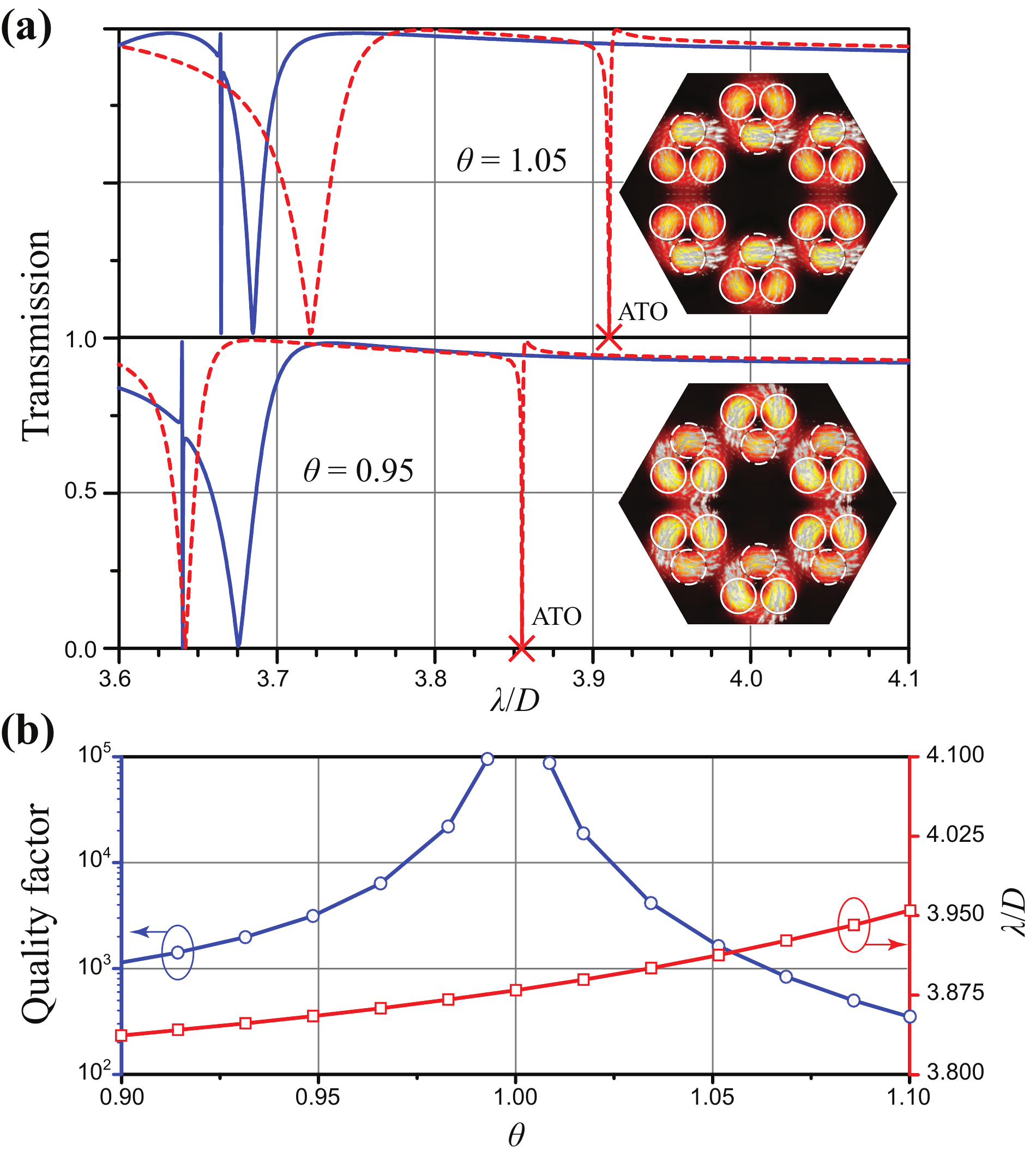}
\caption{(a) Transmitted spectra of the metamaterial with perturbed trimers excited by the $x$-polarized (solid blue lines) and $y$-polarized (dashed red lines) wave. The trimers are perturbed by resonators with different height. The unit supercell symmetry is $C_{2v}$. The normalized magnitude and flow of the magnetic near-field of the ATO mode are given in the inserts, where the perturbed resonators are denoted by dashed circles. (b) Evolution of the quality factor (blue line) and normalized resonant wavelength (red line) of the ATO mode as functions of the asymmetry parameter $\theta$. All parameters of the structure are the same as in Fig.~\ref{fig:fig_7}.}
\label{fig:fig_8}
\end{figure}

High symmetry imposes severe constraints on the possible geometry of the electromagnetic field in the array. The group-theoretical description suggests that the $C_{2v}$ group is the highest symmetry that allows excitation of the ATO mode. This mode can be excited by the wave whose vector ${\bf H}$ is parallel to the $x$ axis ($y$-polarized wave). This selection rule is confirmed by our numerical calculations presented in Fig.~\ref{fig:fig_8}(a), where the resonance occurs in the transmitted spectra directly at the wavelength of the ATO mode. Due to the high symmetry of the unit supercell, the spectral characteristic around the ATO resonance is rather ``clean'', i.e., the toroidal dipole mode appears to be well isolated from other resonances in the spectra, and the wavelength separation of the ATO mode is relatively large. When the vector ${\bf H}$ of the incident wave is parallel to the $y$ axis, none of the trimers is excited, which is in full accordance with the rule of thumb.

The frequency shift and quality factor of the ATO resonance as functions of the asymmetry parameter $\theta$ are presented in Fig.~\ref{fig:fig_8}(b). It is evident that the less the asymmetry, the higher the quality factor, which reaches values higher than $10^5$. The resonant wavelength shifts up as the thickness of the perturbed disks increases.

\subsection{Array with the $C_{s}$ symmetry of unit supercell} %
\begin{figure}[t!]
\centering
\includegraphics[width=\linewidth]{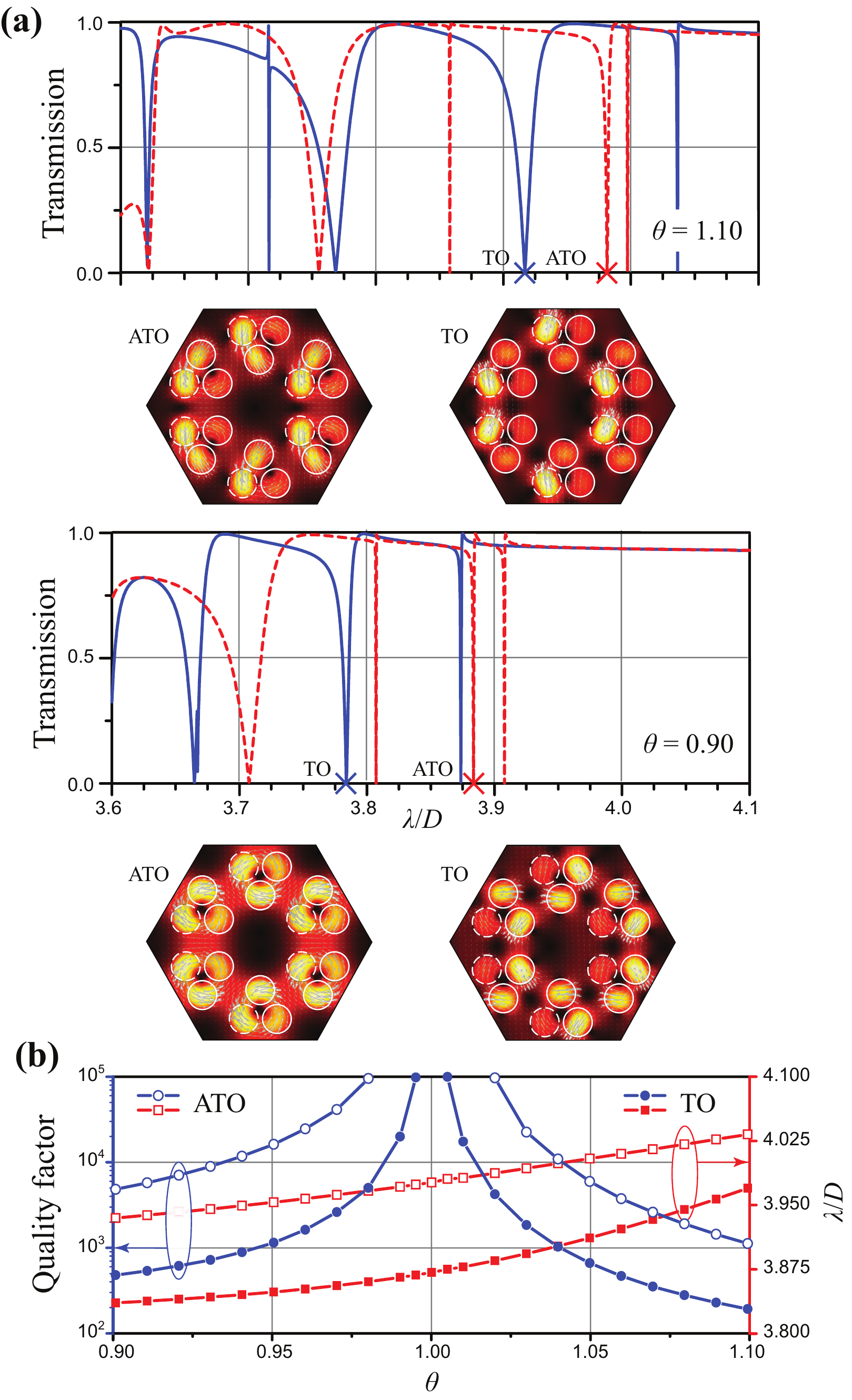}
\caption{Same as in Fig.~\ref{fig:fig_8} but for arrays with the $C_{s}$ symmetry of the hexagonal unit supercell. The corresponding normalized magnitude and flow of the magnetic near-field of the ATO and TO modes are given in the middle planes.}
\label{fig:fig_9}
\end{figure}

The group-theoretical description predicts that in a metamaterial composed of unit supercells with the $C_{s}$ symmetry, both ATO and TO modes can be excited. The results of corresponding calculations are presented in Fig.~\ref{fig:fig_9}(a). These results confirm that the ATO and TO modes arise when the metamaterial is irradiated by the $x$-polarized and $y$-polarized wave, respectively. In general, these resonances occur at different wavelengths. However, due to the low symmetry of the unit supercell, the transmitted spectra are filled with several resonances, so that the ATO and TO resonances appear to be somewhat masked against their background. 

\begin{figure*}[t!]
\centering
\includegraphics[width=\textwidth]{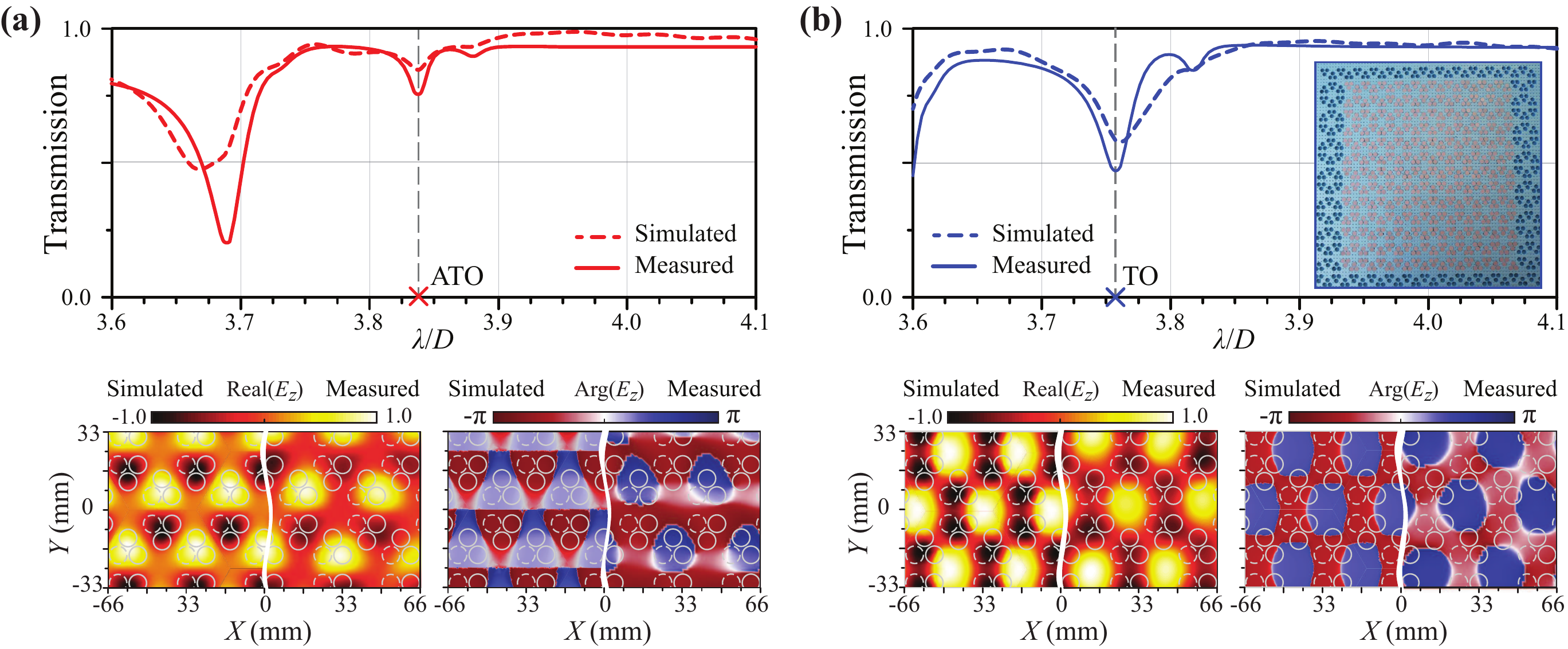}
\caption{Simulated and measured transmitted spectra and resonant patterns of the real part and phase of the $z$-component of the electric near-field for an actual array with the $C_s^{(1)}$ symmetry of the hexagonal unit supercell. The array is irradiated by the (a) $x$-polarized and (b) $y$-polarized wave to excite the ATO and TO mode, respectively. The inset demonstrates a photo of the metamaterial prototype. Parameters of the prototype are: $\varepsilon_d=22\pm 1$, $\tan\delta\approx 1\times 10^{-3}$, $r_d=4$, $h_d=3.5$, $\Delta h_d=-1$ ($\theta \approx 0.71$), $\varepsilon_s=1.1$, $h_s=10$, $a_d=9$, $p=27$. All geometrical parameters are given in millimeters.}
\label{fig:fig_10}
\end{figure*}

Since the material losses are excluded from the present simulations, the maximal value of the quality factor of the given resonances depends only on the asymmetry parameter $\theta$, which defines the coupling degree of the toroidal dipole mode with free space. The net toroidal dipole moment related to the ATO mode is almost zero since the antisymmetric toroidal dipole moments in adjacent trimers compensate each other and reduce the electromagnetic coupling of this mode with free space. The equally oriented moments of the TO mode are lack of this property. Comparing quality factors of the ATO and TO resonances presented in Fig.~\ref{fig:fig_9}(b), one can conclude that the ATO mode possesses the lower radiation losses and consequently higher quality factor than that of the TO mode. In particular, these quality factors differ by one order of magnitude. Besides, the ATO and TO resonances experience the same wavelength shift when the  parameter $\theta$ changes.
\section{Experimental results} 
\label{sec:experiments}
We further validate our theory in neat experiments. To this end, we construct a metamaterial prototype based on the $C_s^{(1)}$ unit supercells since this design provides the efficient coupling of the metamaterial with a linearly polarized incident wave via both the ATO and TO modes. We assemble the array on a dielectric substrate utilizing disks made of a low-loss, high-permittivity microwave ceramic. 

For our experiments, we use a well-established technique that was described in detail earlier \cite{Kupriianov_PIERS_2019}. In particular, we perform our experimental study in the microwave range ($8-15$~GHz) using Rohde \& Schwarz ZVA50 Vector Network Analyzer as the main measurement platform. Our setup also includes a pair of dielectric-lens antennas, a near-field imaging system, and all other necessary accessories. 

The results of our measurements are summarized in Fig.~\ref{fig:fig_10}. It includes a photo of the actual metamaterial prototype, simulated and measured transmitted spectra and electric near-field patterns plotted at the corresponding resonant wavelengths. In the simulations supporting our experimental studies, the real material losses ($\tan\delta$) existing in the metamaterial constituents are taken into account.

Foremost, one can conclude that our experimental data are in good agreement with the simulation results. The measurements performed confirm the existence of polarization-dependent resonances for the ATO and TO modes. Importantly, both resonances survive in a lossy metamaterial. Since the TO mode has a lower quality factor, it can be more easily detected in the experiment.

The calculated and measured characteristics of the electric near-field show a fundamental difference in these two types of toroidal dipole orders. For the ATO mode, the electric near-field is confined in the centers of minicells, while it is practically zero at the center of the hexagonal unit supercells. In the trimers, one can clearly see the staggered distribution of the real part of the normal ($E_z$) component of the electric near-field. This staggered distribution is additionally confirmed by the phase pattern of the $E_z$ component. Contrariwise, for the TO mode, the electric near-field has a bright hotspot in the center of the hexagonal unit supercells complemented by lower intensity hotspots at the center of each minicell. The maximal field concentration is reached outside the dielectric particles, which can be considered as a signature of the toroidal dipole mode. Thus, we have two different conditions of the field concentration in the same metamaterial, where the form of the field concentration is determined only by the given polarization of the incident wave. We believe that this peculiarity is a rather unique feature of the proposed metamaterial, which is very promising for practical applications.
\section{Discussion} 
\label{sec:discussions}
We have focused in this paper only on two possible eigenmodes of the hexagonal unit supercells, namely, the ATO and TO modes which belong to IRREPs $A_1$ and $B_1$ (see  Table~\ref{tab:CharC6v} in Appendix \ref{app:tables}). However, there are also possible resonant modes belonging to other IRREPs, which can be defined using, for example, the method of symmetry adapted linear combinations (SALC) \cite{Dmitriev_2020}.

Symmetry breaking in the unit supercell of the proposed metamaterials can be fulfilled in many different ways. Besides the discussed above methods of the out-of-plane symmetry breaking, the in-plane symmetry breaking can be introduced. For instance, one trimer or a cluster of several trimers in a hexagonal unit supercell can be dislocated or rotated with respect to the other trimers. Effectiveness of the rotation method in the metamaterial composed of square unit cells of trimers was demonstrated recently \cite{Tuz_ACSApplNano_2020}. However, the point symmetry of the resulting unit supercell in this case is usually completely lost, and the symmetry analysis cannot be applied for such perturbed arrays.

In the framework of a chosen symmetry, the geometry is not always defined uniquely due to a high number of disks included in the unit supercell and different kinds of their perturbation. Therefore, the geometry can be optimized following simple rules introduced here. This optimization can be performed before the full-wave numerical simulations, which can significantly reduce the time needed to find the desired configuration of the unit supercell.

The proposed configurations of hexagonal arrays based on trimers possess a technological advantage to have a free space in the center of the hexagonal unit supercell, where an additional element (for example, a nonlinear or control element) can be deposited. In particular, for the case of the excitation of the TO mode, a strong concentration of electric near-field appears in this free space which can be useful in sensoric and lasing applications.

\section{Conclusions} 
\label{sec:conclusions}
We have proposed and studied all-dielectric metamaterials composed of hexagonal trimer-based unit cells which are able to support two types of toroidal supermodes. These supermodes differ in the net order of the toroidal dipole moments. The  moments distributed in trimers can possess either co-directional (symmetric, TO) or staggered (antisymmetric, ATO) arrangement. 

A new theoretical approach based on the magnetic group theory is developed to analyze the mechanism of excitation of these supermodes by the field of a normally incident linearly polarized wave. This mechanism implies symmetry breaking in the supercell of the metamaterial. It is revealed that the ATO mode can be excited in metamaterials whose unit supercells possess the $C_{2v}$, $C_{2}$, and $C_{s}$ symmetries, whereas the TO mode can be excited only in the supercells with the $C_{s}$ symmetry. All these symmetries can be realized by  different arrangements of perturbed trimers in the unit supercell.  

It was demonstrated that the magnetic groups approach simplifies greatly the analysis of the arrays and provides a deeper physical insight into the mechanism of the toroidal modes excitation. 

We have found out that by adjusting the perturbation parameters of the trimers, it is possible to excite the ATO and TO modes in the same metamaterial by the wave with a proper polarization. Excitation of these modes results in different characteristics of the electric near-field localization, which can be important for practical applications.

The method of magnetic groups introduced in this paper can be also applied for the analysis of arrays with square and rectangular unit supercells as well as to the isolated oligomers with high symmetries and their assemblies. In a more wide aspect, this method can be used for the analysis of complex systems with  3D symmetries and eigenmodes different from the toroidal ones.
\section*{Acknowledgment}
VD thanks the Brazilian Agency National Council of Technological and Scientific Development (CNPq) for financial support. SDSS acknowledges support from CNPq (Grant No.~160344/2019-0). ASK and VRT acknowledge financial support from the National Key R\&D Program of China (Project No.~2018YFE0119900). ABE thanks funding support from the Deutsche Forschungsgemeinschaft (DFG, German Research Foundation) under Germany’s Excellence Strategy within the Cluster of Excellence PhoenixD (EXC 2122, Project ID 390833453).

\appendix
\section{\label{app:Schoenflies}Elements of point symmetry in Sch\"oenflies notation \cite{Bradley_book_2009}}

In Sch\"oenflies system, an $n$-fold rotation through $2\pi/n$ (where $n$ is an integer; in our case $n=2,3,6$)  about the $z$ axis is denoted by the symbol $C_n$ ($C$ means Cyklus). The symbol $\sigma_v$ (the subscript $v$ for vertical) defines reflection in a plane passing through this axis. The symbol $\sigma_d$ (the subscript $d$ for diagonal) designates a mirror plane containing the axis which is diagonal to the already existing plane $\sigma_v$. 

Let us apply now to the group notations. The groups with one axis of symmetry are denoted by $C_n$. Notice that in this case, the notations for the group elements and groups themselves coincide. For example, the symbol $C_2$ denotes the operation of rotation about an axis by $\pi$, and also it may denote the $C_2$ group consisting of two elements: the identity $e$ and the rotation $C_2$. The meaning of the notations can be clearly understood from the context. The $C_{nv}$ group has an $n$-fold rotational axis $C_n$ and a finite number of planes of symmetry passing through the axis $C_n$. The $C_s$ group contains two elements: the identity $e$ and the reflection~$\sigma_v$.

\begin{figure}[htbp]
\centering\includegraphics[width=30mm]{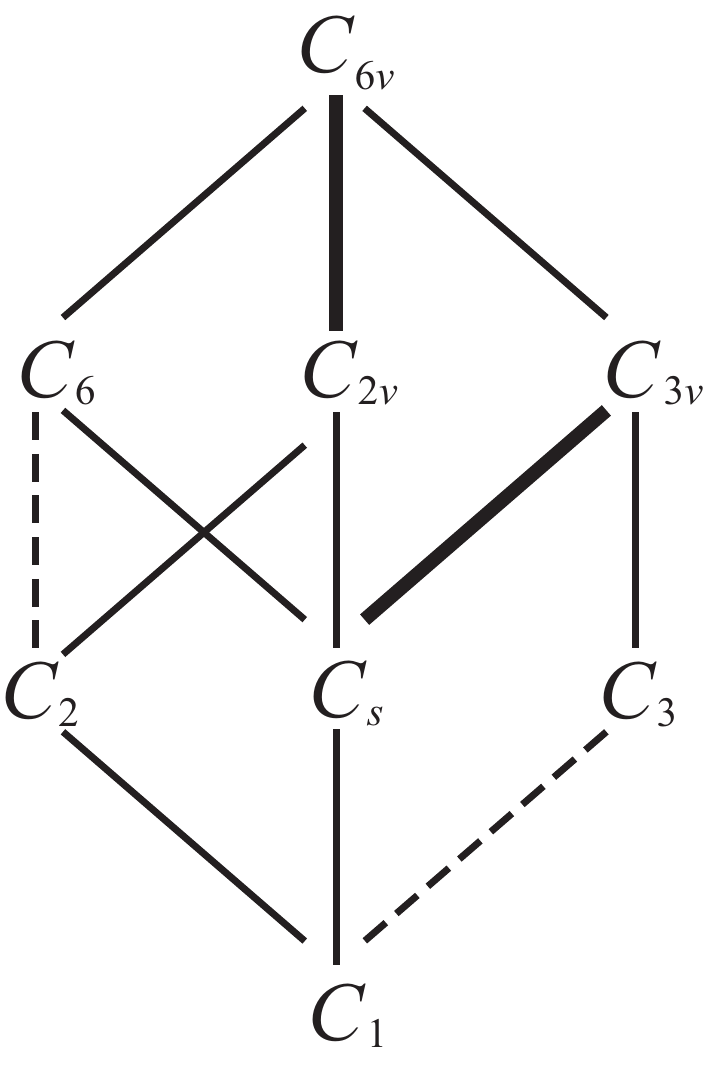}
\caption{Group decomposition of  $C_{6v}$ group. Thick and dotted lines indicate that the corresponding subgroup is not invariant and is not of index 2 with respect to the higher group, respectively. It means that these two groups cannot form a group of the third category \cite{Barybin_book_2002}.}
\label{fig:fig_A1}
\end{figure}

The subgroup decomposition (the group tree) of the $C_{6v}$ group, which is discussed in the main body of this paper, is shown in Fig.~\ref {fig:fig_A1}. The $C_{6v}$ group has the following subgroups: $C_{6}$, $C_{3v}$, $C_{3}$, $C_{2v}$, $C_{2}$, $C_{s}$, and $C_{1}$.

\section{\label{app:tables}Tables of group theory}

\begin{table}[H]
\begin{center}
\caption{Character table of  $C_{6v}$ group and magnetic groups of the second and third categories, and corresponding possible toroidal dipole mode orders.}
    \begin{tabular}{c@{\hspace{1mm}}c@{\hspace{1mm}}c@{\hspace{1mm}} 
    c@{\hspace{1mm}}c@{\hspace{1mm}}c@{\hspace{1mm}}c@{\hspace{1mm}}c@{\hspace{1mm}}	c@{\hspace{1mm}}}\hline \hline
	\fbox{$C_{6v}$}   & $~e~~$   & $C_2$ & 2$C_3$ & 2$C_6$ & $3\sigma_v$ & $3\sigma_d$ & Magnetic group & Mode order\\   \hline \hline
	$A_1$ & 1   & 1   & 1  &  1  &  1  & 1  &  $\bf C_{6v}$ &  dark TO\\
	$A_2$ & 1  & 1 & 1 & 1  &$\!\!\!\!\! -1$  &$\!\!\!\!\! -1$  &  $C_{6v}(C_6)$    \\
	$B_1$ & 1   &$\!\!\!\!\! -1$ & 1   &$\!\!\!\!\! -1$   & 1   &$\!\!\!\!\! -1$  &   $C_{6v}(C_{3v})$ & dark ATO\\
	$B_2$ & 1     &$\!\!\!\!\! -1$   & 1     &$\!\!\!\!\! -1$    &$\!\!\!\!\! -1$  & 1 & $C_{6v}(C_{3v})$ \\
	$E_1$ & 2     &$\!\!\!\!\! -2$   &$\!\!\!\!\! -1$ & 1     &$0$    & $0$  & \\
	$E_1$ & 2     &$2$   &$\!\!\!\!\! -1$ & $\!\!\!\!\! -1$    &$0$    & $0$  & 
	\\ \hline \hline
    \end{tabular}
\label{tab:CharC6v}
\end{center}
\end{table}
\begin{table}[H]
\begin{center}
\caption{Irreducible representations of  $C_{2v}$ group and  their relation to magnetic groups, and corresponding possible toroidal dipole mode orders.}
		{
		\begin{tabular}{c@{\hspace{2mm}}c@{\hspace{2mm}}c@{\hspace{2mm}} 
				c@{\hspace{2mm}}c@{\hspace{2mm}}c@{\hspace{2mm}}c	
				}\hline \hline
				\fbox{$C_{2v}$}   & $~e~$   & $C_2$  & $\sigma_{1}$ &
				$\sigma_{2}$  & Magnetic group & Mode order \\  \hline\hline
				$A_1$ & 1   & 1   & 1  &  1  &  $\bf C_{2v}$ & dark TO \\
				$A_2$ & 1  & 1   &$\!\!\!\!\! -1$  &$\!\!\!\!\! -1$  &   $C_{2v}(C_2)$ &   \\
				$B_1$ & 1   &$\!\!\!\!\! -1$ & 1   &$\!\!\!\!\! -1$     
			  &  $C_{2v}(C_{s})$ & bright ATO  \\
				$B_2$ & 1     &$\!\!\!\!\! -1$       &$\!\!\!\!\! -1$  & 1
				& $C_{2v}(C_{s})$ &
				\\\hline \hline
			\end{tabular}
		}
		\label{tab:ERREPC2v}
	\end{center}
\end{table}
\begin{table}[H]
\begin{center}
\caption{Irreducible representations of  $C_{s}$ group and their relation to magnetic groups of the second and third categories, and corresponding possible toroidal dipole mode orders. Superscripts in $C_s$ denote orientation of the $\sigma$ plane in the hexagonal cluster, [see Figs. \ref{fig:fig_4}(g) and \ref{fig:fig_4}(h)].}
\begin{tabular}{c@{\hspace{2mm}}c@{\hspace{2mm}}c@{\hspace{2mm}} 
 c@{\hspace{2mm}}c@{\hspace{2mm}}c@{\hspace{2mm}}c	
	}  	\hline \hline		
\fbox{$C_{s}$}  & $~e~$ & $\sigma_v$ &  Magnetic group & Mode order\\ \hline 
\hline
 $A$ & $1$ & $1$  & ${\bf C}_s^{(1)}$ & bright TO \\
 $B$ & $1$ & $\!\!\!\!-1$ & $C_s^{(1)}(C_1)$ & bright ATO \\ \hline 
 \hline
\end{tabular}
\label{tab:IRREPCs}
	\end{center}
	\end{table}
\begin{table}[H]
\begin{center}
\caption{Symmetry degeneration table of  $C_{6v}$ group into  $C_{2v}$ and $C_{s}$ groups \cite{Overvig_PhysRevB_2020}. Superscripts in $C_s$ denote orientation of the $\sigma$ plane in the hexagonal cluster [see Figs. \ref{fig:fig_4}(g) and \ref{fig:fig_4}(h)].} 
\begin{tabular}{c@{\hspace{2mm}}c@{\hspace{2mm}}c@{\hspace{2mm}} 
c@{\hspace{2mm}}c@{\hspace{2mm}}c@{\hspace{2mm}}c	
				}\hline \hline			
$C_{6v}$  & $C_{2v}$        & $C_{s}^{(1)}$       & $C_{s}^{(2)}$\\ \hline  \hline
				$A_1$ & $A_1$         & $A$          & $A$  
				\\
				$A_2$ & $A_2$         &  $B$         &  $B$ 
				\\
				$B_1$ & $B_1$         & $A$          & $B$ 
				\\
				$B_2$ & $B_2$         &  $B$         &  $A$ 
				\\ 
				$E_1$ & $B_1$, $B_2$  & $A, B$       & $A,B$  
				\\
				$E_2$ & $A_1$, $A_2$  &  $A, B$      &  $A,B$ 
					 			\\ 
					 		\hline	\hline
								\end{tabular}
\label{tab:1}	
		\label{tab:degeneration}
	\end{center}
\end{table}

\section{\label{app:magnetic}Brief description of magnetic groups}

{\it Time reversal operator}. The time reversal operator ${\cal T}$ can be an element of magnetic groups entering in these groups either separately or in combination with elements of the geometrical symmetry. The ${\cal T}$ operator changes the sign of time $t$ ($t\rightarrow-t$), commutes with all elements of the geometrical symmetry, and has the property ${\cal T} {\cal T}={\cal T}^2=e$, where $e$ is the unit element of the group.

Here we shall discuss only those properties of the ${\cal T}$ operator which are necessary for our present consideration. In particular, the ${\cal T}$ operator reverses the velocities and changes the current directions, signs of magnetic fluxes, magnetic fields, toroidal moments, wave vector, and Poynting vector. All these quantities are odd in time. 

{\it Categories of magnetic groups}. There exist three categories of discrete and continuous point magnetic groups. The group of the first category $G$ consists of a unitary subgroup $H$ (in our case, it contains the usual rotation-reflection elements) and products of ${\cal T}$ with all the elements of $H$. The full group is then $H+{\cal T}H$ including~${\cal T}={\cal T}e$ (these groups are also referred to as nonmagnetic ones, see the first column of Table~\ref{tab:content}).

In the case of magnetic groups of the second category ${\bf G}$, there is no point group elements combined with the time reversal operator ${\cal T}$, and ${\cal T}$ itself is not an element of the groups. The nomenclature and notations of the groups of the first (nonmagnetic) and second (magnetic) category coincides. Therefore, to distinguish them, we  use bold letters for denoting the groups of the second category (see  second column of Table~\ref{tab:content}).
%
\begin{table}[htbp]
	\begin{center}
			\caption{Content of magnetic groups of symmetry}
{
\begin{tabular}{@{}c@{}c@{}c@{}c@{}}%
  \hline \hline 
1st category & \quad 2nd category & 3rd category         \\
\hline \hline   $ G=H+{\cal T}H$      & $\bf G$      &
$G(H)=H+{\cal T}H^{\prime},  H^{\prime}\neq H$\\ 
including ${\cal T}$  & without ${\cal T}$  &  ${\cal T}$ only in combination    \\
               &              & with rotation-reflections  \\
\hline\hline 
\end{tabular}
} 
\label{tab:content}
\end {center}
\end{table}

The Sch\"oenflies system is particularly suitable for notation of the magnetic groups of the third category. In this case, the notation presents explicitly the structure of the group, i.e., the unitary subgroup (it contains only elements of the geometrical symmetry) and the antiunitary elements (these elements are the product of the operator ${\cal T}$ and an element of the geometrical symmetry).

In addition to the rotation-reflection elements of the subgroup $H$ of pure geometrical operators, the  magnetic groups of the third category $G(H)$ also contain the elements which are the product of ${\cal T}$ and the usual geometrical symmetry elements. These combined elements cause the existence of the so-called antiaxes and antiplanes of symmetry. The full group is $H+{\cal T}H^{\prime}$ without ${\cal T}$. Notice that the elements of $H^{\prime}$ are different from those of $H$ (see  third column of Table~\ref{tab:content}).

Geometrical elements of a magnetic group of the third category form a subgroup of index 2. It means that in each group of the third category, there is an equal number of elements with and without ${\cal T}$ (see Fig.~\ref{fig:fig_A1}). In contrast to the groups of the first category, the operator ${\cal T}$ itself is not an element of the magnetic groups of the third category.

\section{\label{app:dynamic}``Dynamic'' magnetic symmetry of alternating magnetic field, magnetic, and toroidal dipole moments}
The term {\it dynamic symmetry} is used in classical and quantum mechanical systems when describing dynamical interactions and revealing the relations between dynamics and geometry. Description of the dynamic symmetry is based on the Lie groups. In particular, in electromagnetic theory, the dynamic symmetry is related to the invariance of the Maxwell equations under the conformal group of transformations \cite{Wulfman_2011}. In our present consideration, the notion of ``dynamic'' symmetry has another meaning, which is discussed below. That is why we put the term {\it dynamic} in quotes.

The theory of nonmagnetic groups is frequently used to describe optical properties of metamaterials without magnetic inclusions (e.g., see Ref. \cite{Padilla_OptExpress_2007}), while a traditional application of magnetic groups is related to the description of systems with a static magnetic field and a static magnetization in magnetic substances. An external static magnetic field together with geometrical symmetry of the medium constituents defines symmetry of the whole magnetic system. This is used, for instance, to calculate the composition of permittivity and permeability tensors (i.e., to impose some restrictions on the elements of these tensors) and also to define the structure of the scattering matrices \cite{Barybin_book_2002}.

In our case, we shall consider two alternating magnetic fields with different symmetries. They are the fields of the incident wave and toroidal dipole mode. Although these magnetic fields interact in a nonmagnetic environment of the array of dielectric particles, an approach based on the magnetic groups makes it possible to get a deep physical insight into the problem of the toroidal mode excitation.

\begin{figure}[htbp]
\centering
\includegraphics[width=75mm]{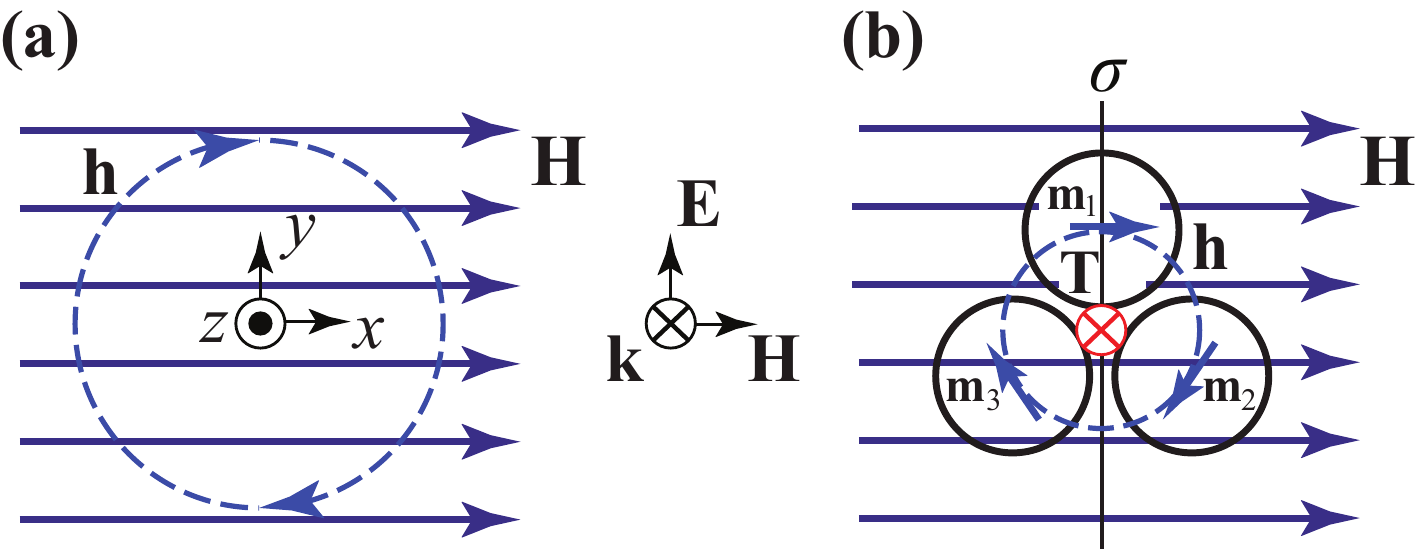}
\caption{(a) Electric $\bf E$ and magnetic $\bf H$ fields of an incident wave with the wave vector $\bf k$ and circular magnetic field $\bf h$ of an idealized toroidal dipole mode. (b) Orientation of the magnetic dipole moments ${\bf m}_i$ ($i=1,2,3$) and toroidal dipole moment $\bf T$ in the magnetic field $\bf H$ of the incident wave for the trimer with plane of symmetry $\sigma$.}
\label{fig:fig_D1}
\end{figure}

Let us consider a trimer based on an equilateral triangle illuminated by a normally incident linearly polarized wave with the wave vector $\bf k$. For the chosen geometry of the problem, the magnetic field of the incident wave $\bf H$ lies in the $x$-$y$ plane. We suppose that a toroidal dipole mode is excited in the trimer. This mode appears from the circular flow of the magnetic field ${\bf h}$, which also lies in the $x$-$y$ plane. The appearance of the vectors $\bf H$ and $\bf h$ is schematically given in Fig.~\ref{fig:fig_D1}.

To relate the symmetries of the magnetic fields of the incident wave $\bf H$ and toroidal dipole mode $\bf h$, two prerequisites must be fulfilled: (i) the loop of the magnetic field ${\bf h}$ is uniform along its circumference, and (ii) a phase shift between the fields $\bf H$ and ${\bf h}$ is absent. It means that these two dynamic quantities are in phase being considered at a fixed point of time. This time can be fixed at an arbitrary moment, for example, when the fields reach their maximal values during the oscillation period.

In nonmagnetic symmetry, the magnetic field $\bf H$ has only one element of symmetry, namely, the plane perpendicular to the vector $\bf H$. In magnetic symmetry, the magnetic field $\bf H$ can be described by the 2D (in the $x$-$y$ plane) magnetic symmetry, namely, by the group of the third category $C_{2v}(C_s)$, which contains the following elements and antielements:

$\bullet$\, the unit element $e$, 

$\bullet$\, the antiaxis ${\cal T}C_{2}$ along the $z$ axis,

$\bullet$\, the vertical plane $\sigma_1$ perpendicular to the vector $\bf H$ (the plane $x=0$), 

$\bullet$\, the vertical antiplane  ${\cal T}\sigma_2$, parallel to the vector $\bf H$ (the plane $y=0$).\\
In this group, the time reversal operator ${\cal T}$ is combined with the geometrical elements of the $C_{2}$ symmetry and~$\sigma_2$.

In an ideal case, the magnetic field $\bf h$ of the toroidal mode presents a circle [see Fig.~\ref{fig:fig_D1}(a)]. Considering the $x$-$y$ plane where this circle is situated, the 2D group of symmetry of $\bf h$ is $\mbox{\boldmath {\bf C}}_{\infty v}$ of the second category with the $z$ axis of infinite order ${C}_{\infty}$, and an infinite number of the planes of symmetry $\sigma_v$ passing through this axis. The time reversal operator ${\cal T}$ is not the element of $\mbox{\boldmath {\bf C}}_{\infty v}$ since when $t\rightarrow-t$, the sign of $\bf h$ changes.

Comparing symmetries of the fields, one can see that the fields $\bf H$ and $\bf h$ have the antiaxis  ${\cal T}C_{2}$ and axis $C_{2}$, respectively. Moreover, the field $\bf H$ has the antiplane ${\cal T}\sigma$, while the field $\bf h$ has the plane $\sigma$ which coincides with ${\cal T}\sigma$. ${\cal T}C_{2}$ and $C_{2}$, and also ${\cal T}\sigma$ and $\sigma$  are incompatible with each other. This incompatibility can be considered as a selection rule. Therefore, in the given symmetry,  excitation of the toroidal dipole mode having a circular flow of $\bf h$ induced by the field $\bf H$ is impossible. 

Now we apply to a realistic case. In an isolated trimer, there is a particular eigensolution where three in-plane magnetic dipoles ${\bf m}_i$ ($i=1,2,3$) appears to be arranged in a head-to-tail fashion [Fig.~\ref{fig:fig_D1}(b)]. In the linear regime, the moment $\bf m$ of the magnetic dipole in a dielectric disk is proportional to the incident magnetic field $\bf H$: 
\begin{equation}
{\bf m} =\alpha \bf H,
\label{eq:magnetic}
\end{equation}
where $\alpha$ is the magnetic polarizability. The trimer based on an equilateral triangle can be described by symmetry $C_{3v}$ with the $C_3$ axis directed along the $z$ axis and three planes of symmetry $\sigma$. Evidently, the cluster of three magnetic dipoles ${\bf m}_i$ inherits the magnetic symmetry  $\mbox{\boldmath {\bf C}}_{3v}$. In this case, the excitation of the toroidal dipole mode in the trimer by the incident field $\bf H$ is also impossible \cite{Dmitriev_2020}. 

The excitation of the toroidal dipole mode in the trimer becomes possible after removing the $C_3$ axis and two planes of its symmetry $\sigma$. It can be made by introducing some perturbation to the trimer geometry (e.g., by changing the geometric dimensions or material parameters of one disk in the trimer \cite{Dmitriev_2020}). This perturbation reduces the geometric symmetry of the trimer to the $C_s$ group.  Notice that the position of the perturbed disk in the trimer defines the orientation of its symmetry plane $\sigma$, and, consequently, imposes a constraint on the orientation of the vector $\bf H$ necessary for the toroidal dipole mode excitation.

\bigskip

\bibliography{Hexagon}

\end{document}